\documentclass[aps,prl,superscriptaddress,reprint,twocolumn]{revtex4-1}
\usepackage{xspace}
\usepackage{graphicx}
\usepackage{color}
\usepackage{amsmath, amsthm, amssymb}
\usepackage{natbib}

\newcommand{\ucsd}{Department of Physics, University of California,
                          San Diego, La Jolla CA 92093}
\newcommand{\ricebioen}{Department of Bioengineering, Rice University, Houston, TX}
\newcommand{\ricectbp}{Center for Theoretical Biological Physics, Rice University, Houston, TX}

\newcommand{\rot}{\ensuremath{\mathcal{R}}}
\newcommand{\ri}{\ensuremath{\vb{r}}^i\xspace}
\newcommand{\dri}{\ensuremath{\boldsymbol\delta \vb{r}^i}\xspace}
\newcommand{\qi}{\ensuremath{\vb{q}}^i\xspace}
\newcommand{\betai}{\ensuremath{\beta^i\xspace}}
\newcommand{\betab}{\ensuremath{\bar{\beta}\xspace}}
\newcommand{\taub}{\ensuremath{\bar{\tau}\xspace}}

\newcommand{\rj}{\ensuremath{\vb{r}}^j\xspace}
\newcommand{\rij}{\ensuremath{\hat{\vb{r}}^{ij}}\xspace}
\newcommand{\dij}{\ensuremath{d^{ij}}\xspace}
\newcommand{\poi}{\ensuremath{\vb{p}}^i\xspace}
\newcommand{\eq}{Eq.~}

\newcommand{\fig}{Fig.~}

\newcommand{\vb}[1]{ {\mathbf #1}}

\newcommand{\rb}{\ensuremath{\vb{r}}\xspace}

\newcommand{\pmt}[4]{\ensuremath{\begin{pmatrix} #1 & #2 \\ #3 & #4 \end{pmatrix}}}

\newcommand{\meanc}[1]{\ensuremath{\langle{#1}\rangle_c}\xspace}

\newcommand{\ci}{\ensuremath{\textrm{CI}}\xspace}

\begin{document}
\title{Emergent collective chemotaxis without single-cell gradient sensing}
\author{Brian~A.~Camley}
\affiliation{\ucsd}
\author{Juliane~Zimmermann}
\affiliation{\ricectbp}
\author{Herbert~Levine}
\affiliation{\ricectbp}\affiliation{\ricebioen}
\author{Wouter-Jan~Rappel}
\affiliation{\ucsd}
\begin{abstract}
Many eukaryotic cells chemotax, sensing and following chemical gradients. However, experiments have shown that even under conditions when single cells cannot chemotax, small clusters may still follow a gradient. This behavior has been observed in neural crest cells, in lymphocytes, and during border cell migration in Drosophila, but its origin remains puzzling.  Here, we propose a new mechanism underlying this ``collective guidance", and study a model based on this mechanism both analytically and computationally.  Our approach posits that the contact inhibition of locomotion (CIL), where cells polarize away from cell-cell contact, is regulated by the chemoattractant.  Individual cells must measure the mean attractant value, but need not measure its gradient, to give rise to directional motility for a cell cluster. We present analytic formulas for how cluster velocity and chemotactic index depend on the number and organization of cells in the cluster. The presence of strong orientation effects provides a simple test for our theory of collective guidance.
\end{abstract}

\maketitle

Cells often perform chemotaxis, detecting and moving toward increasing concentrations of a chemoattractant, to find nutrients or reach a targeted location.  This is a fundamental aspect of biological processes from immune response to development.  Many single eukaryotic cells sense gradients by measuring how a chemoattractant varies over their length \cite{levine2013physics}; this is distinct from bacteria that measure chemoattractant over time \cite{segall1986temporal}.  In both, single cells are capable of net motion toward higher chemoattractant.  

Recent measurements of how neural crest cells respond to the chemoattractant Sdf1 suggest that single neural crest cells cannot chemotax effectively, but small clusters can \cite{theveneau2010collective}. A more recent report shows that at low gradients, clusters of lymphocytes also chemotax without corresponding single cell directional behavior; at higher gradients clusters actually move in the opposite direction to single cells \cite{malet2015collective}.  In addition, late border cell migration in the {\it Drosophila} egg chamber may occur by a similar mechanism \cite{bianco2007two,rorth2007collective,inaki2012effective,wang2010light}.  These experiments strongly suggest that gradient sensing in a cluster of cells may be an {\em emergent} property of cell-cell interactions, rather than arising from amplifying a single cell's biased motion; interestingly, some fish schools also display emergent gradient sensing \cite{berdahl2013emergent}.  In fact, these experiments led to a ``collective guidance" hypothesis \cite{rorth2007collective}, in which a cluster of cells where each individual cell has no information about the gradient may nevertheless move directionally.  In a sense that will become clear, cell-cell interactions allow for a measurement of the gradient across the entire cluster, as opposed to across a single cell.

In this paper, we develop a quantitative model that embodies the collective guidance hypothesis.  Our model is based on modulation of the well-known contact inhibition of locomotion (CIL) interaction~\cite{lin2015interplay,carmona2008contact,desai2013contact}, in which cells move away from neighboring cells. We propose that individual cells measure the local signal concentration and adjust their CIL strength accordingly; the cluster moves directionally due to the spatial bias in the cell-cell interaction. We discuss the suitability of this approach for explaining current experiments, and provide experimental criteria to distinguish between chemotaxis via collective guidance and other mechanisms where clusters could gain improvement over single-cell migration \cite{simons2004many,coburn2013tactile}. These results may have relevance to collective cancer motility \cite{friedl2012classifying}, as recent data suggest that tumor cell clusters are particularly effective metastatic agents \cite{aceto2014circulating}.  

\begin{figure}[ht!]
\centering
\includegraphics[width=85mm]{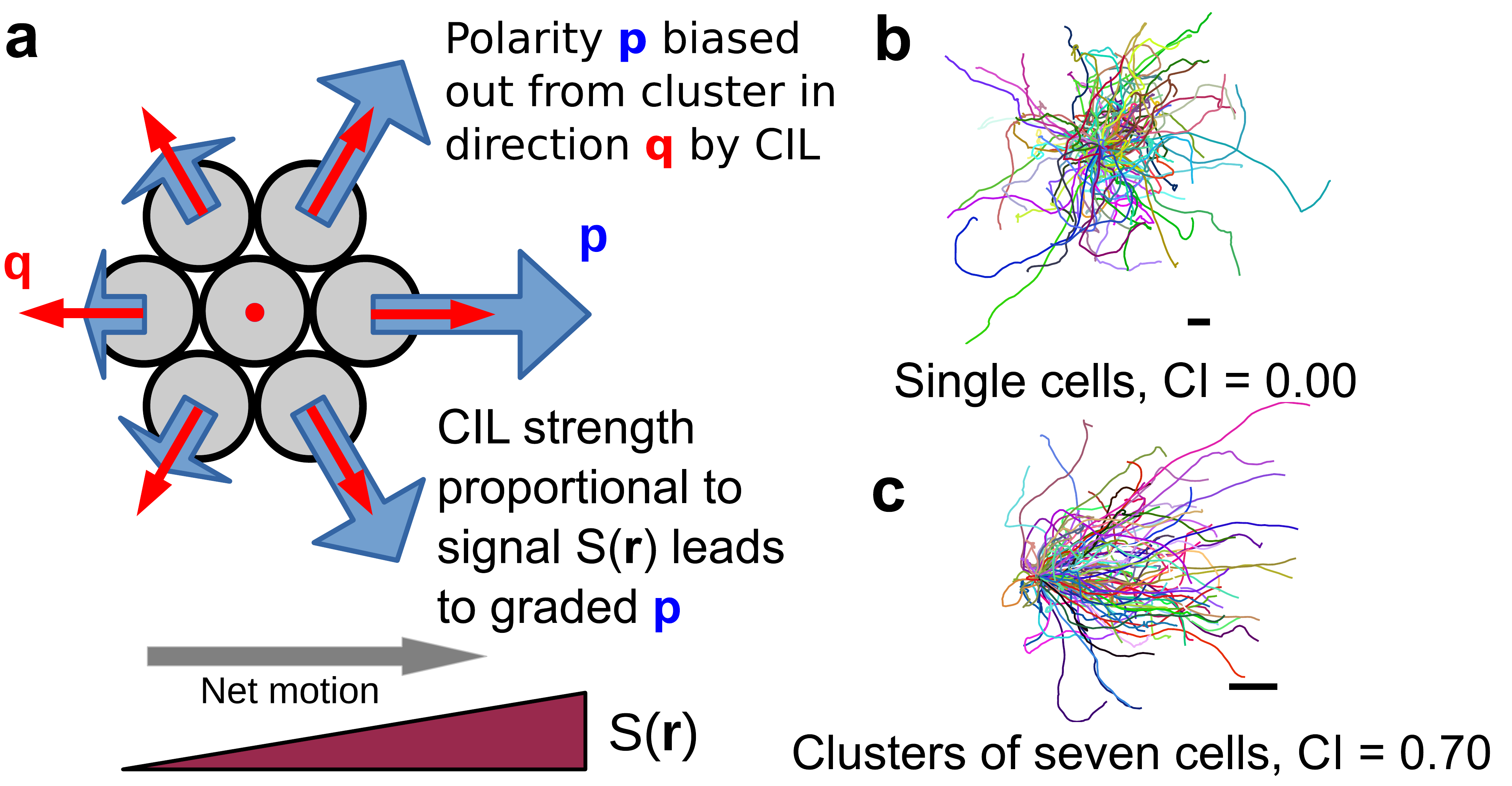}
\caption{\linespread{1.0}\selectfont{}{\bf Signal-dependent contact inhibition of locomotion creates directed motion}.  {\bf a,} Schematic picture of model and origin of directed motion.  Cell polarities are biased away from the cluster toward the direction $\qi = \sum_{j \sim i} \rij$ by contact inhibition of locomotion (CIL); the strength of this bias is proportional to the local chemoattractant value $S(\rb)$, leading to cells being more polarized at higher $S$.  See text for details.    {\bf b,} One hundred trajectories of a single cell and {\bf c,} cluster of seven cells.  Trajectories are six persistence times in length (120 min).  Scalebar is one cell diameter.  The gradient strength is $|\nabla S| = 0.025$ in these simulations, with the gradient in the $x$ direction.}  
\label{fig:schematic}
\end{figure}

We consider a cluster of cells exposed to a chemical gradient $S(\rb)$.   
We use a two-dimensional stochastic particle model to describe cells, 
giving each cell $i$ a position $\ri$ and a polarity $\poi$.  The cell polarity indicates its direction and propulsion strength: an isolated cell with polarity $\poi$ has velocity $\poi$.  The cell's motion is overdamped, so the velocity of the cell is $\poi$ plus the total physical force other cells exert on it, $\sum_{j \neq i} \vb{F}^{ij}$.  Biochemical interaction between cells alter a cell's polarity $\poi$.  Our model is then:
\begin{align}
\label{eq:position} \partial_t \ri &= \poi + \sum_{j \neq i} \vb{F}^{ij} \\
\label{eq:polarity} \partial_t \poi &= -\frac{1}{\tau} \poi + \sigma {\boldsymbol\xi}^i(t) + \betai \sum_{j \sim i}\rij 
\end{align}
where $\vb{F}^{ij}$ are the intercellular forces of cell-cell adhesion and volume exclusion, and ${\boldsymbol\xi}^{i}(t)$ are Gaussian Langevin noises with $\langle \xi^i_\mu(t) \xi^j_\nu(t') \rangle = 2 \delta_{\mu\nu} \delta^{ij} \delta(t-t')$, where the Greek indices $\mu,\nu$ run over the dimensions $x,y$.  The first two terms on the right of \eq \ref{eq:polarity} are a standard Ornstein-Uhlenbeck model \cite{selmeczi2005cell,vankampen}: $\poi$ relaxes to zero with a timescale $\tau$, but is driven away from zero by the noise $\boldsymbol\xi(t)$.  This corresponds with a cell that is orientationally persistent over a time of $\tau$.  

We have introduced the last term on the right of \eq \ref{eq:polarity} to describe contact inhibition of locomotion (CIL).  CIL is a well-known property of many cell types in which cells polarize away from cell-cell contact \cite{carmona2008contact,mayor2010keeping,camley2014polarity,abercrombie1979contact,desai2013contact}.  We model CIL by biasing $\poi$ away from nearby cells, toward $\qi = \sum_{j \sim i} \rij$, where $\rij = (\ri - \rj) / |\ri-\rj|$ is the unit vector pointing from cell $j$ to cell $i$ and the sum over $j \sim i$ indicates the sum over the neighbors of $i$ (those cells within a distance $D_0 = 1.2$ cell diameters).  While this is motivated by CIL in neural crest, it is also a natural minimal model under the assumption that cells know nothing about their neighbors other than their direction $\rij$.   For cells along the cluster edge, the CIL bias $\qi$ points outward from the cluster, but for interior cells $\qi$ is smaller or zero (\fig \ref{fig:schematic}a).  This is consistent with experimental observations that edge cells have a strong outward polarity, while interior cells have weaker protrusions \cite{theveneau2010collective}.

Chemotaxis arises in our model if the chemoattractant $S(\rb)$ changes a cell's susceptibility to CIL, $\betai$, $\betai = \betab S(\rb^i)$.  This models the result of \cite{theveneau2010collective} that the chemoattractant Sdf1 stabilizes protrusions induced by CIL \cite{theveneau2010collective}.  We also assume that the cell's chemotactic receptors are not close to saturation - i.e. the response is perfectly linear.  If CIL is present even in the absence of chemoattractant ($S = 0$), as in neural crest  \cite{theveneau2010collective}, i.e. $\betai = \beta_0 + \betab S(\rb)$, this will not significantly change our analysis.  Similar results can also be obtained if all protrusions are stabilized by Sdf1 ($\tau$ regulated by $S$), though with some complications ({\it Appendix}, \fig A1). 

{\it Analytic predictions for cluster velocity.--}Our model predicts that while single cells do not chemotax, clusters as small as two cells will, consistent with \cite{theveneau2010collective}.  We can analytically predict the mean drift of a cluster of cells obeying Eqs. \ref{eq:position}-\ref{eq:polarity}:
\begin{equation}
\langle \vb{V} \rangle_c \approx \betab \tau \mathcal{M}\cdot \nabla S \label{eq:shallow}
\end{equation}
where the approximation is true for shallow gradients, $S(\rb) \approx S_0 + \rb\cdot \nabla S$.  $\meanc{\cdots}$ indicates an average over the fluctuating $\poi$ but with a fixed configuration of cells $\ri$.  The matrix $\mathcal{M}$ only depends on the cells' configuration,
\begin{equation}
\mathcal{M}_{\mu\nu} = \frac{1}{N}\sum_{i} q^i_\mu r^i_\nu \label{eq:matrix}
\end{equation}
where, as above, $\qi = \sum_{j \sim i} \rij$.  \eq \ref{eq:shallow} resembles the equation of motion for an arbitrarily shaped object in a low Reynolds number fluid under a constant force $\betab\tau \nabla S$ \cite{kim2013microhydrodynamics}: by analogy, we call $\mathcal{M}$ the ``mobility matrix."  There is, however, no fluctuation-dissipation relationship as there would be in equilibrium \cite{han2006brownian}.  

To derive \eq \ref{eq:shallow}, we note that \eq \ref{eq:position} states that, in our units, the velocity of a single cell is equal to the force on it (i.e. the mobility is one).  For a cluster of $N$ cells, the mean velocity of the cluster is $1/N$ times the total force on the cluster.  As $\vb{F}^{ij} = - \vb{F}^{ji}$, the cluster velocity is $\vb{V} = N^{-1} \sum_{i} \poi$.  When the cluster configuration changes slowly over the timescale $\tau$, \eq \ref{eq:polarity} can be treated as an Ornstein-Uhlenbeck equation with an effectively time-independent bias from CIL.  The mean polarity is then $\langle \poi \rangle = \betai \tau \sum_{j \sim i} \rij$, with Gaussian fluctuations away from the mean, $\langle (\poi_\mu -\langle\poi_\mu\rangle)^2 \rangle = \sigma^2 \tau$. The mean cell cluster velocity is
\begin{equation}
\meanc{\vb{V}} = \frac{\betab \tau}{N} \sum_{i} S(\ri) \sum_{j \sim i} \rij
\end{equation}
In a constant chemoattractant field, $S = S_0$, no net motion is observed, as $\sum_{i} \sum_{j \sim i} \rij = 0$.  For linear or slowly-varying gradients $S(\rb) \approx S_0 + \rb \cdot \nabla S$, and we get \eq \ref{eq:shallow}.  

\begin{figure}[ht!]
\includegraphics[width=90mm]{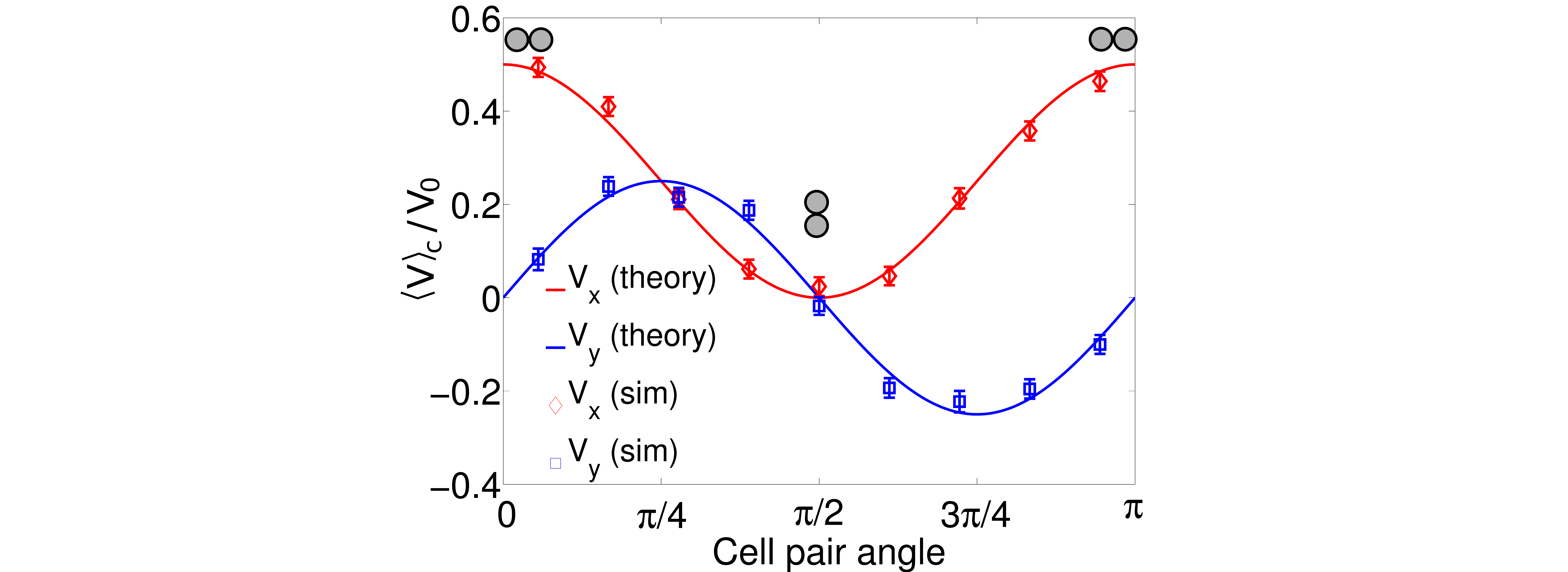}
\caption{\linespread{1.0}\selectfont{}{\bf Adherent pairs of cells undergo highly anisotropic chemotaxis.}  The average chemotactic velocity of a highly adherent cell pair $\meanc{V_x}$ depends strongly on the angle $\theta$ between the cell-cell axis and the chemotactic gradient.  Cell pairs also drift perpendicular to the gradient, $\meanc{V_y}\neq 0$.  $V_0 \equiv \betab\tau |\nabla S|$ is the velocity scale; $|\nabla S| = 0.025$.  Simulations are of Eqs. \ref{eq:position}-\ref{eq:polarity}.  We compute $\meanc{V_\mu}$ by tracking the instantaneous angle, then averaging over all velocities within the appropriate angle bin.
Error bars here and throughout are one standard deviation of the mean, calculated from a bootstrap. Over $n = 13,000$ trajectories of 6$\tau$ (120 minutes) are simulated.}
\label{fig:pairs}
\end{figure}

{\it Cluster motion and chemotactic efficiency depend on cluster size, shape, and orientation.--} Within our model, a cluster's motion can be highly anisotropic.  Consider a pair of cells separated by unit distance along $(\cos \theta,\sin\theta)$.  Then by \eq \ref{eq:matrix}, $\mathcal{M}_{xx} = \frac{1}{2}\cos^2\theta$, $\mathcal{M}_{xy} = \mathcal{M}_{yx} = \frac{1}{2}\cos\theta\sin\theta$, $\mathcal{M}_{yy} = \frac{1}{2} \sin^2\theta$.  If the gradient is in the $x$ direction, then $\meanc{V_x} = \frac{V_0}{2} \cos^2\theta$ and $\meanc{V_y} = \frac{V_0}{2} \cos \theta \sin\theta$, where $V_0 = \betab\tau|\nabla S|$.  Cell pairs move toward higher chemoattractant, but their motion is along the pair axis, leading to a transient bias in the $y$ direction before the cell pair reorients due to fluctuations in $\poi$ (\fig \ref{fig:pairs}).  We compare our theory for the motility of rigid cell clusters (\eq \ref{eq:shallow}) with a simulation of \eq \ref{eq:position}-\ref{eq:polarity} with strongly adherent cell pairs with excellent agreement (\fig \ref{fig:pairs}).  

{For the simulations in \fig \ref{fig:pairs} and throughout the paper, we solve the model equations Eqs. \ref{eq:position}-\ref{eq:polarity} numerically using a standard Euler-Maruyama scheme.  We choose units such that the equilibrium cell-cell separation (roughly 20 $\mu$m for neural crest \cite{theveneau2010collective}) is unity, and the relaxation time $\tau = 1$ (we estimate $\tau = 20$ minutes in neural crest \cite{theveneau2010collective}). Within these units, neural crest cell velocities are on the order of $1$, so we choose $\sigma = 1$ -- this corresponds to a root mean square speed of an isolated cell being $\langle |\vb{V}|^2 \rangle^{1/2} = 2^{1/2} \sigma \tau^{1/2} \approx 1.4$ microns/minute.  The typical cluster velocity scale is $V_0 = \betab \tau |\nabla S|$, which is 0.5 (0.5 microns/minute in physical units) if $|\nabla S| = 0.025$ and $S(0) = 1$, corresponding to $\beta^i$ changing by 2.5\% across a single cell at the origin.  Cell-cell forces $\vb{F}^{ij}$ are chosen to be stiff springs so that clusters are effectively rigid (see {\it Appendix} for details).  }

We can also compute $\mathcal{M}$ and hence $\langle V_x \rangle$ for larger clusters (Table S1, {\it Appendix}, \fig A2).  For a cluster of $Q$ layers of cells surrounding a center cell, $\mathcal{M}_{\mu\nu} = f(Q) \delta_{\mu\nu}$, with $f(Q) = \frac{  9 Q^2 + 3 Q}{2+6 Q + 6 Q^2}$.  A cluster with $Q$ layers has $N = 1 + 3Q + 3Q^2$ cells; thus the mean velocity of a $Q$-layer cluster is given by $\langle V_x \rangle/V_0 = \overline{\mathcal{M}} = \frac{ 3 N - \sqrt{12 N - 3}}{2 N}$, where $\overline{\mathcal{M}} = \frac{1}{2} \left(\mathcal{M}_{xx} + \mathcal{M}_{yy}\right)$ is the angular average of $\mathcal{M}$.  We predict that $\langle V_x \rangle/V_0$ first increases with $N$, then slowly saturates to $3/2$.  This is confirmed by simulations of the full model (\fig \ref{fig:clustersize}a).  We note that $\langle V_x \rangle$ is an average over time, and hence orientation (see below, {\it Appendix}).  We can see why $\langle V_x \rangle$ saturates as $N \to \infty$ by considering a large circular cluster of radius $R$.  Here, we expect $\qi = a \hat{\vb{n}}$ on the outside edge, where $a$ is a geometric prefactor and $\hat{\vb{n}}$ is the outward normal, with $\qi = \vb{0}$ elsewhere. Then, $\mathcal{M}_{\mu\nu} \sim \frac{a}{\pi R^2}\int_{0}^{2\pi} \left( R d\theta \right) \hat{\vb{n}}_\mu(\theta) \vb{r}_\nu = 2a \delta_{\mu\nu}$, independent of cluster radius $R$.    
A related result has been found for circular clusters by Malet-Engra et al. \cite{malet2015collective}; we note that they do not consider the behavior of single cells or cluster geometry.  

The efficiency of cluster chemotaxis may be measured by chemotactic index (\ci), commonly defined as the ratio of distance traveled along the gradient (the $x$ displacement) to total distance traveled \cite{fuller2010external}; \ci ranges from -1 to 1.  We define $\ci \equiv \langle V_x \rangle / \langle |\vb{V}| \rangle$, where the average is over both time and trajectories (and hence over orientation).  The chemotactic index \ci may also be computed analytically, and it depends on the variance of $\vb{V}$, which is $\langle \left(V_x-\langle V_{x} \rangle \right)^2 \rangle = \langle \left(V_y-\langle V_{y} \rangle \right)^2 \rangle = \sigma^2 \tau / N$.  In our model, $\ci$ only depends on the ratio $c$ of mean chemotactic velocity to its standard deviation,
\begin{align}
\nonumber \ci &= \sqrt{2/\pi} c / L_{1/2}(-c^2/2)\\
c &= \frac{\langle V_x \rangle}{\sqrt{\langle \left(V_\mu-\langle V_{\mu} \rangle \right)^2 \rangle}} = \frac{\betab \tau  \overline{\mathcal{M}} |\nabla S|}{\sigma \sqrt{\tau/N}}
\label{eq:ci}
\end{align}
where $L_{1/2}$ is a generalized Laguerre polynomial.  When mean cluster velocity is much larger than its fluctuations, $c \gg 1$ and $\ci\to1$, but when fluctuations are large, $|c| \ll 1$ and $\ci \to 0$ ({\it Appendix}, \fig A3).  Together, \eq \ref{eq:shallow}, \eq \ref{eq:ci} and Table S1 provide an analytic prediction for cluster velocity and \ci, with excellent agreement with simulations (\fig \ref{fig:clustersize}).  We note that $\langle V_x \rangle/V_0$ only depends on cluster configuration, where $V_0 = \betab\tau|\nabla S|$, so $\langle V_x(N) \rangle/V_0$ collapses onto a single curve as the gradient strength is changed (\fig \ref{fig:clustersize}a).  By contrast, how \ci increases with $N$ depends on $|\nabla S|$ and $\sigma$ (\eq \ref{eq:ci}, \fig \ref{fig:clustersize}b).

\begin{figure}[ht!]
\includegraphics[width=85mm]{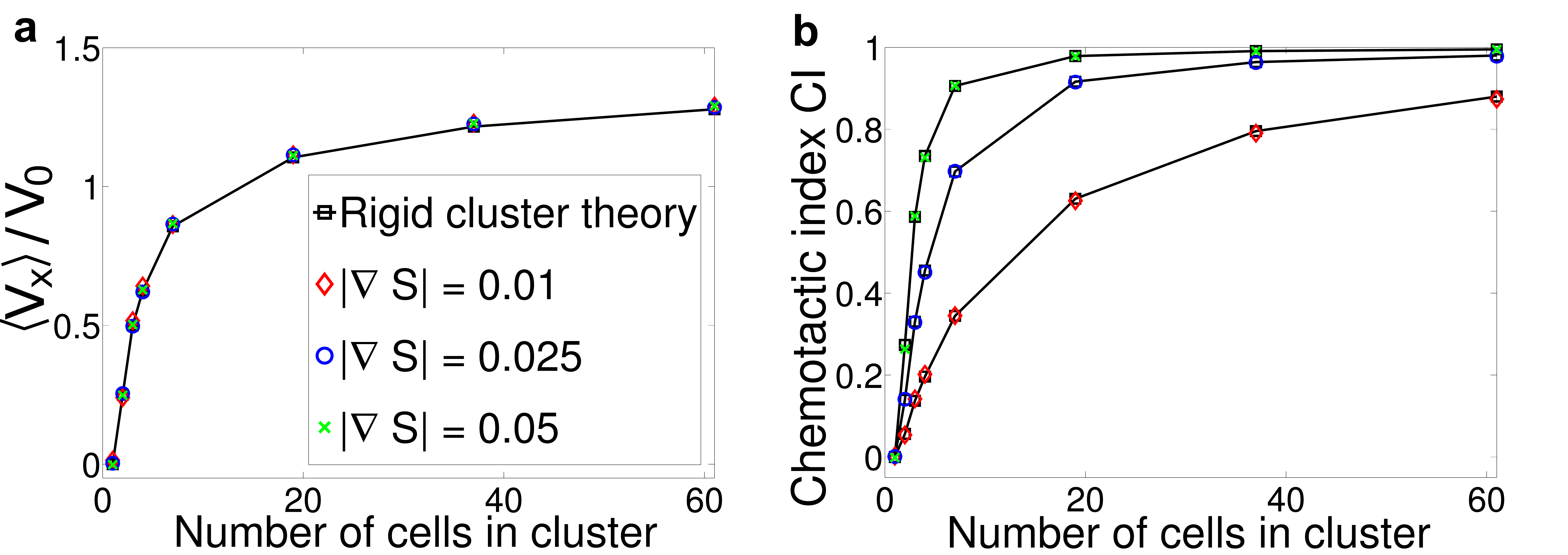}
\caption{\linespread{1.0}\selectfont{}{\bf Larger cell clusters chemotax more effectively, but their velocity saturates}  {\bf a,} As the number of cells $N$ in a cluster increases, the mean velocity $\langle V_x \rangle$ increases with $N$ but then saturates; the mean velocity can be collapsed onto a single curve by rescaling by $V_0 \equiv \betab\tau |\nabla S|$.  {\bf b,} The chemotactic index \ci also saturates to its maximum value.  Black squares and lines are the orientationally-averaged drift velocity computed for rigid clusters by \eq \ref{eq:shallow} and \eq \ref{eq:ci}.  Colored symbols are full model simulations with strong adhesion.  Cell cluster shape may influence $\langle V_x \rangle$ ({\em Appendix} \fig A4); our calculations are for the shapes in Table S1.  Error bars here are symbol size or smaller; $n \ge 2000$ trajectories of $6\tau$ are used for each point.}
\label{fig:clustersize}
\end{figure}

In our model, clusters can in principle develop a spontaneous rotation, but in practice this effect is small, and absent for symmetric clusters (see {\it Appendix}).  

{{\it Motion in non-rigid clusters.--}  While we studied near-rigid clusters above, our results hold qualitatively for clusters that are loosely adherent and may rearrange.  Cell rearrangements are common in many collective cell motions \cite{angelini2010cell,angelini2011glass,szabo2010collective,vedula2013collective}, but we note that in \cite{malet2015collective} clusters are more rigid.  We choose cell-cell forces $\vb{F}^{ij}$ to allow clusters to rearrange (see {\it Appendix}, \cite{warren2003vapor}), and simulate Eqs. \ref{eq:position}-\ref{eq:polarity}.  As in rigid clusters, $\langle V_x \rangle$ increases and saturates, while \ci increases toward unity, though more slowly than a rigid cluster (\fig \ref{fig:fluid}ab).  Clusters may fragment; with increasing $x$, $\beta^i$ increases and the cluster breaks up (\fig \ref{fig:fluid}c).  Cluster breakup can limit guidance -- if $\betab$ is too large, clusters are not stable, and will not chemotax.}

{In \fig \ref{fig:fluid}ab, we compute \ci and velocity by averaging over all cells, not merely those that are connected.  If we track cells ejected from the cluster, they have an apparent $\ci > 0$, as they are preferentially ejected from the high-$\beta^i$ cluster edge ({\it Appendix}).  Experimental analysis of dissociating clusters may therefore not be straightforward.  Anisotropic chemotaxis is present in non-rigid pairs, though lessened because our non-rigid pairs rotate quickly with respect to $\tau$ ({\it Appendix}).  
\begin{figure}[ht!]
\centering
\includegraphics[width=90mm]{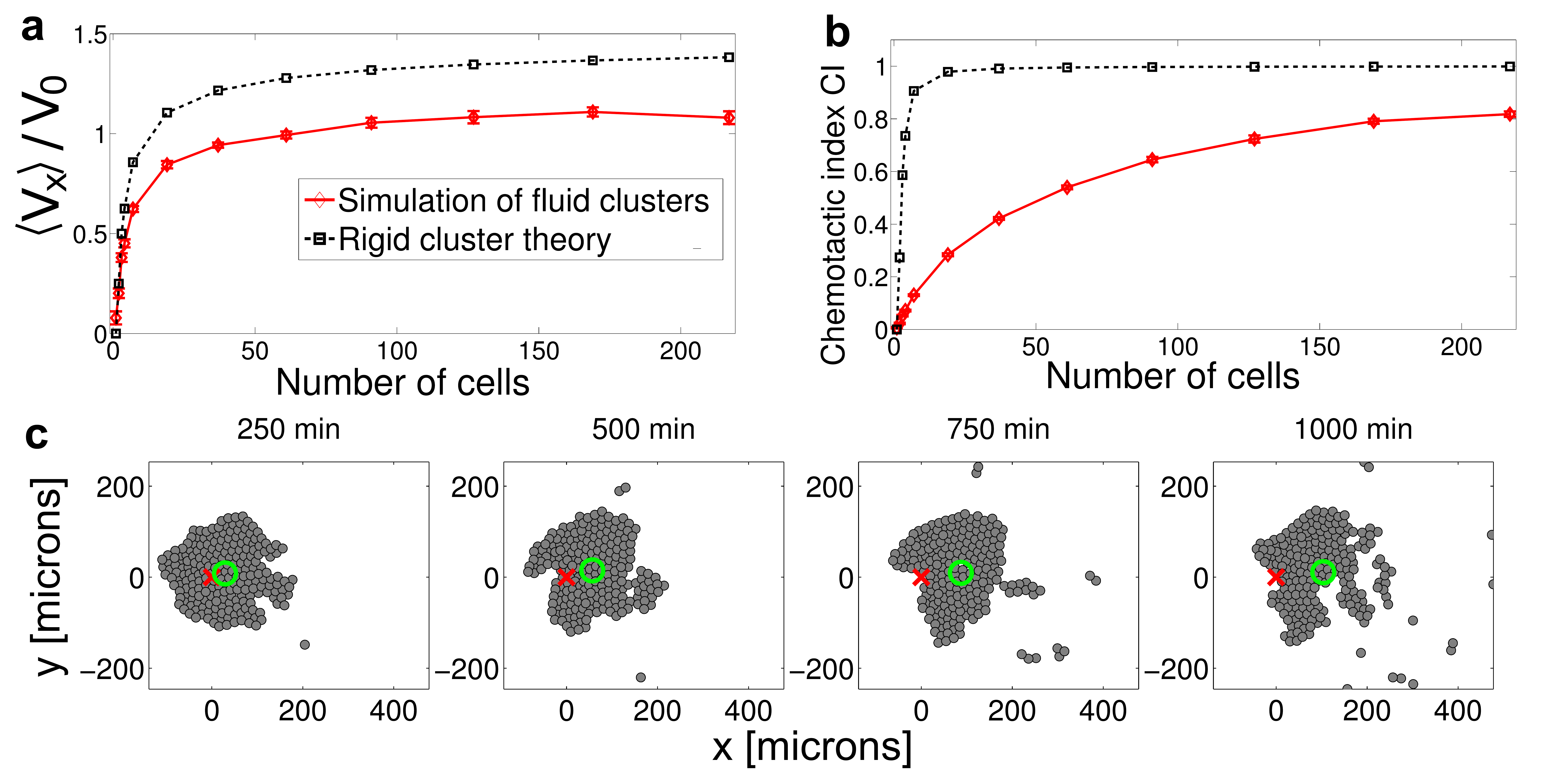}
\caption{\linespread{1.0}\selectfont{} {\bf Nonrigid clusters may also chemotax via collective guidance}.  {\bf a,} As the number of cells $N$ in a cluster increases, the mean velocity $\langle V_x \rangle$ increases with $N$ but then saturates.  {\bf b,} Chemotactic index also approaches unity, but slower than in a rigid cluster.  Rigid cluster theory assumes the same cluster geometries as in \fig \ref{fig:clustersize}. Averages in {\bf a-b} are over $n \ge 20$ trajectories (ranging from $n = 20$ for $N = 217$ to $n = 4000$ for $N = 1,2$), over the time $12.5\tau$ to $50\tau$.  {\bf c,} Breakdown of a cluster as it moves up the chemoattractant gradient.  X marks the initial cluster center of mass, O the current center. $|\nabla S| = 0.1$, $\betab = 1$ in this simulation.} 
\label{fig:fluid}
\end{figure}}

{\it Distinguishing between potential collective chemotaxis models.--}Our model explains how chemotaxis can emerge from interactions of non-chemotaxing cells.  However, other possibilities exist for enhancement of chemotaxis in clusters.  Coburn et al. showed that in contact-based models, a few chemotactic cells can direct many non-chemotactic ones \cite{coburn2013tactile}.  If single cells are weakly chemotactic, cell-cell interactions could amplify this response or average out fluctuations \cite{simons2004many}.  How can we distinguish these options?  In lymphocytes \cite{malet2015collective}, the motion of single cells oppositely to the cluster immediately rules out simple averaging or amplification of single cell bias. More generally, the scaling of collective chemotaxis with cluster size does not allow easy discrimination. In \fig \ref{fig:clustersize}, at large $N$, $\langle V_x \rangle$ and \ci saturate.   As an alternate theory, suppose each cell chemotaxes noisily, e.g. $\poi = p_0 \nabla S + \boldsymbol\Delta^i$, where $\boldsymbol\Delta$ are independent zero-mean noises.  In this case, $\langle \vb{V} \rangle = p_0 \nabla S$ independent of $N$, and $\langle (V_\mu - \langle V_\mu \rangle)^2 \rangle \sim 1/N$, as in our large-$N$ asymptotic results and the related circular-cluster theory of \cite{malet2015collective}.  Instead,  we propose that orientation effects in small clusters are a good test of emergent chemotaxis.  In particular, studying cell pairs as in \fig \ref{fig:pairs} is critical: anisotropic chemotaxis is a generic sign of cluster-level gradient sensing.  Even beyond our model, chemotactic drift is anisotropic for almost all mechanisms where single cells do not chemotax, because two cells separated perpendicular to the gradient sense the same concentration.  This leads to anisotropic chemotaxis unless cells integrate information over times much larger than the pair's reorientation time. By contrast, the simple model with single cell chemotaxis above leads to isotropic chemotaxis of pairs.

How well does our model fit current experiments? We find increasing cluster size increases cluster velocity and chemotactic index.  This is consistent with \cite{malet2015collective}, who see a large increase in taxis from small clusters ($<20$ cells) to large, but not \cite{theveneau2010collective}, who find that \ci is similar between small and large clusters, and note no large variations in velocity.  {This suggests that the minimal version of collective guidance as developed here can create chemotaxis, but does not fully explain the experiments of \cite{theveneau2010collective}.}  There are a number of directions for improvement.  More quantitative comparisons could be made by detailed measurement of single-cell statistics \cite{selmeczi2005cell,amselem2012stochastic}, leading to nonlinear or anisotropic terms in \eq \ref{eq:polarity}.  Our description of CIL has also assumed, for simplicity, that both cell front and back are inhibitory; other possibilities may alter collective cell motion \cite{camley2014polarity}.  We could also add adaptation as in the LEGI model \cite{levchenko2002models,takeda2012incoherent} to enable clusters to adapt their response to a value independent of the mean chemoattractant concentration.
We will treat extensions of this model elsewhere; our focus here is on the simplest possible results.

In summary, we provide a simple, quantitative model that embodies a minimal version of the collective guidance hypothesis \cite{rorth2007collective,theveneau2010collective} and provides a plausible initial model for collective chemotaxis when single cells do not chemotax.  Our work allows us to make an unambiguous and testable prediction for emergent collective guidance: pairs of cells will develop anisotropic chemotaxis. Although there has been considerable effort devoted to models of collective motility \cite{sepulveda2013collective,camley2014velocity,szabo2006phase,li2014coherent,czirok1996formation,van2014collective,basan2013alignment,zimmermann2014intercellular,szabo2010collective,segerer2015emergence}, ours is the first model of how collective chemotaxis can emerge from single non-gradient-sensing cells via collective guidance and regulation of CIL. 

\begin{acknowledgments}
BAC appreciates helpful discussions with Albert Bae and Monica Skoge.  This work was supported by NIH Grant No. P01 GM078586, NSF Grant No. DMS 1309542, and by the Center for Theoretical Biological Physics.  BAC was supported by NIH Grant No. F32GM110983. 
\end{acknowledgments}

\onecolumngrid
\section{Appendix}
\setcounter{equation}{0}
\setcounter{figure}{0}
\setcounter{table}{0}
\renewcommand*{\thefigure}{A\arabic{figure}}
\renewcommand*{\thetable}{A\arabic{table}}
\renewcommand*{\theequation}{A\arabic{equation}}

\section{Cluster chemotaxis when chemoattractant regulates cell persistence $\tau$}


Within the main paper, we have assumed that the chemoattractant concentration $S(\rb)$ regulates the susceptibility of a cell to contact inhibition of locomotion $\beta^i$, with $\betai = \betab S(\ri)$.  This models the stabilization of protrusions induced by contact interactions.  This is consistent with the results of Theveneau et al. \cite{theveneau2010collective}, who find that protrusion stabilization is stronger in clusters than in single cells.  However, very similar results can be found if we assume that $\beta$ is constant and the signal regulates the time required for the cell's polarity to relax, i.e. $\tau^i = \taub S(\rb)$.  In this case, the mean polarity of a cell is $\langle \poi \rangle = \beta \tau^i \sum_{j \sim i} \rij$ and we find
\begin{equation}
\langle \vb{V} \rangle \approx \beta \taub \mathcal{M}\cdot \nabla S \; \; (\tau \,\textrm{ regulation})
\end{equation}
where the mobility matrix $\mathcal{M}$ is the same as in the main paper,
$\mathcal{M}_{\mu\nu} = \frac{1}{N}\sum_{i} r^i_\nu q^i_\mu $.
However, because $\tau$ varies over space, the fluctuations will also vary: $\langle \left(V_\mu-\langle V_{\mu} \rangle \right)^2 \rangle = \sigma^2 N^{-2} \sum_i \tau^i = \sigma^2 N^{-1} \taub \overline{S}$, where $\overline{S} = N^{-1} \sum_i S(\ri)$ is the mean signal across the cluster.  For this reason, the chemotactic index in the $\tau$-regulation model will depend on $\nabla S / \overline{S}^{1/2}$, and will not be constant over a linear gradient.  

In addition, a single cell with a persistence time $\tau$ that depends on the chemoattractant level will undergo biased motion.  This is shown in \fig \ref{fig:singletau} below.  This drift can be made smaller than the CIL-driven cluster drift, as it is independent of $\beta$, while the cluster drift is proportional to $\beta$.  

\begin{figure}[ht!]
\begin{centering}
\includegraphics[width=80mm]{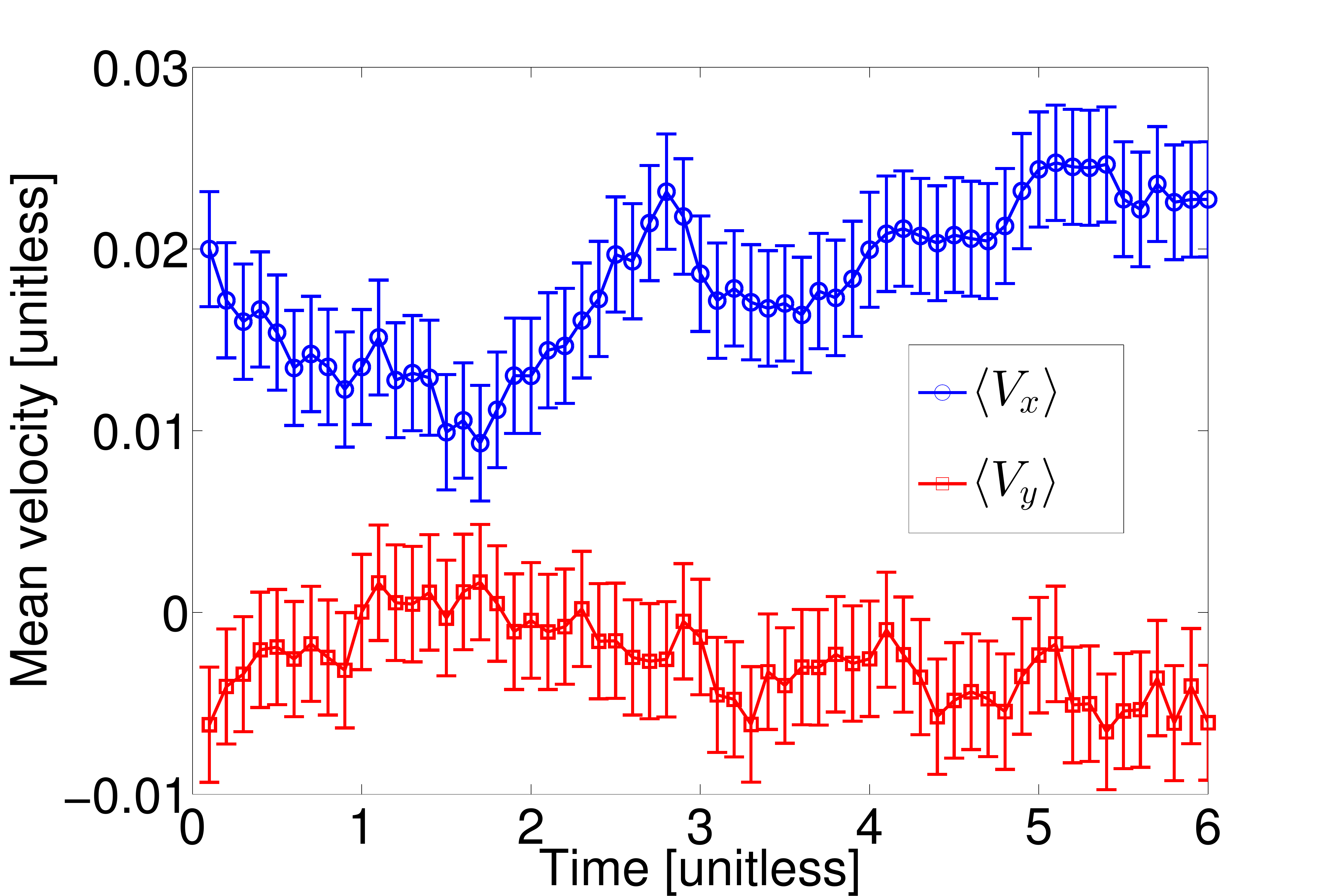}
\caption{{\bf Single cells in a spatially-varying $\tau$ develop a mean drift}. The mean $x$ and $y$ velocities for a cell with spatially varying $\tau$ are shown: $\tau = \overline{\tau}\left(S_0 + |\nabla S| x \right)$, with $\overline{\tau}=1$, $S_0 = 1$, $|\nabla S| = 0.025$.  Result is average over $n=10^5$ iterations, each started at the origin; error bars indicate $\langle \left[V_\mu(t)-\langle V_\mu(t) \rangle\right]^2 \rangle^{1/2}/\sqrt{n}$.  }
\label{fig:singletau}
\end{centering}
\end{figure}


\section{Derivation of the mobility matrices for $Q$-layer oligomers}

\begin{figure}[ht!]
\begin{centering}
\includegraphics[width=80mm]{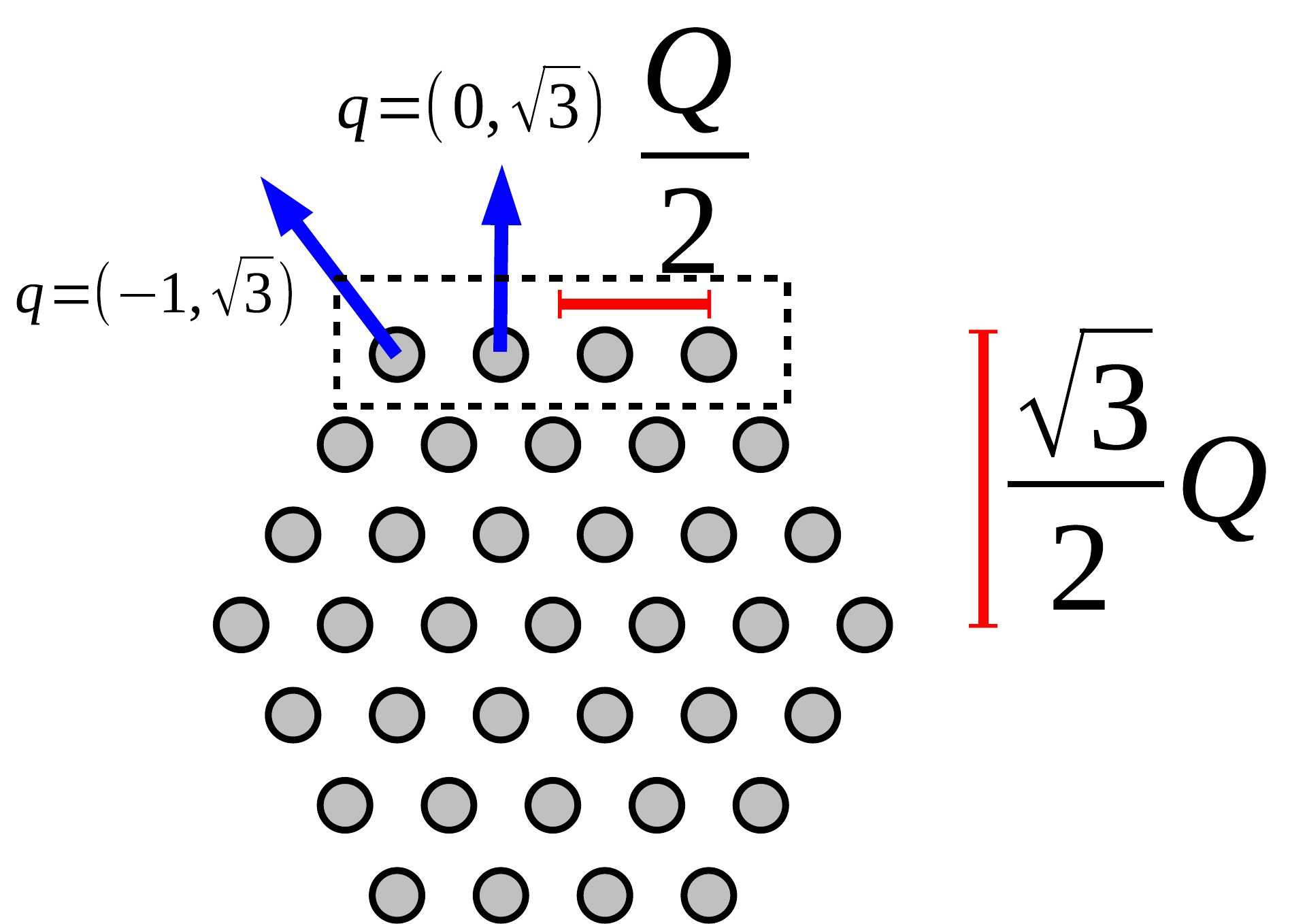}
\caption{{\bf Geometry of $Q$-layer oligomer, illustrated for $Q=3$}.  The top face is highlighted by a dashed line.  }
\label{fig:qmer}
\end{centering}
\end{figure}

We can compute the mobility matrix of the $Q$-layer oligomers for arbitrary $Q$.  Our mobility matrix is given by \eq \ref{eq:matrix}
with $\qi = \sum_{j \sim i} \rij$.  To simplify the calculation, we can make a few assumptions.  First, we note that $\mathcal{M}_{xx} = \mathcal{M}_{yy}$, but $\mathcal{M}_{xy} = \mathcal{M}_{yx} = 0$ for the $Q$-layer oligomer.  We only need to calculate $\overline{\mathcal{M}} = \frac{1}{2} \left( \mathcal{M}_{xx} + \mathcal{M}_{yy} \right)$.  The only cells $i$ in the sum of \eq \ref{eq:matrix} that are nonzero are those around the boundary.  $\overline{\mathcal{M}}$ does not depend on orientation, so we can compute the sum $\sum_i \vb{r}^i \vb{q}^i$ for one face of the oligomer (\fig \ref{fig:qmer}), then multiply by six.  However, this double-counts the corner cells, so we must weight them by $1/2$.  We then find $\overline{\mathcal{M}} = \left[\frac{9}{2} Q^2 + \frac{3}{2} Q \right]/N(Q)$, where $N(Q) = 1 + 3 Q + 3 Q^2$ is the number of cells in the cluster.  

We present the mobility matrices for both $Q$-layer oligomers and other cluster shapes in Table \ref{tab:mobs}.  

\begin{table}[h]
\def\arraystretch{1.4}
\centering
\begin{tabular}{|c|c|c|}
\hline 
{\bf Shape} & $\mathcal{M}$ & Angularly averaged $\overline{\mathcal{M}}$ \\
\hline 
\begin{tabular}{c} Dimer \\ \includegraphics[width=0.6cm]{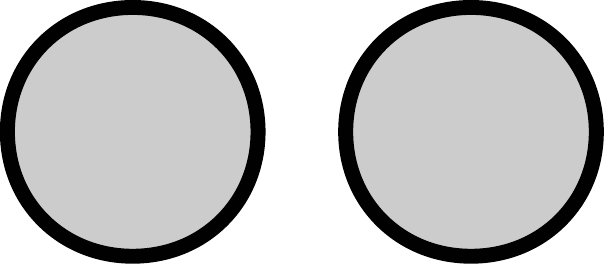} \end{tabular} 
& \pmt{1/2}{0}{0}{0}  & 1/4 \\
\begin{tabular}{c} Trimer \\ \includegraphics[width=0.6cm]{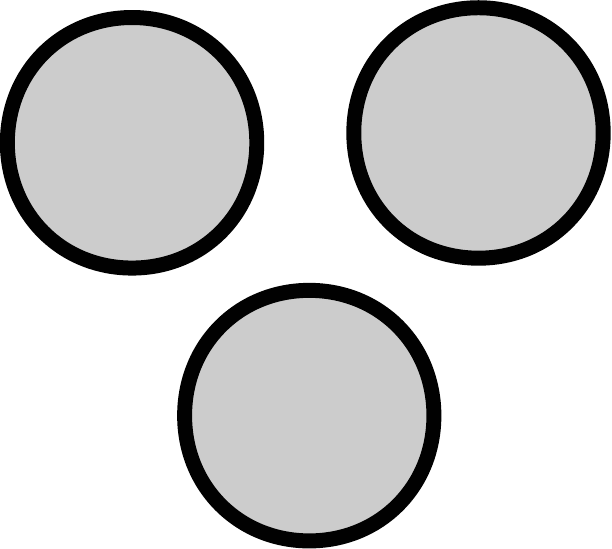} \end{tabular}  & \pmt{1/2}{0}{0}{1/2}  & 1/2 \\
\begin{tabular}{c} Tetramer \\ \includegraphics[width=0.6cm]{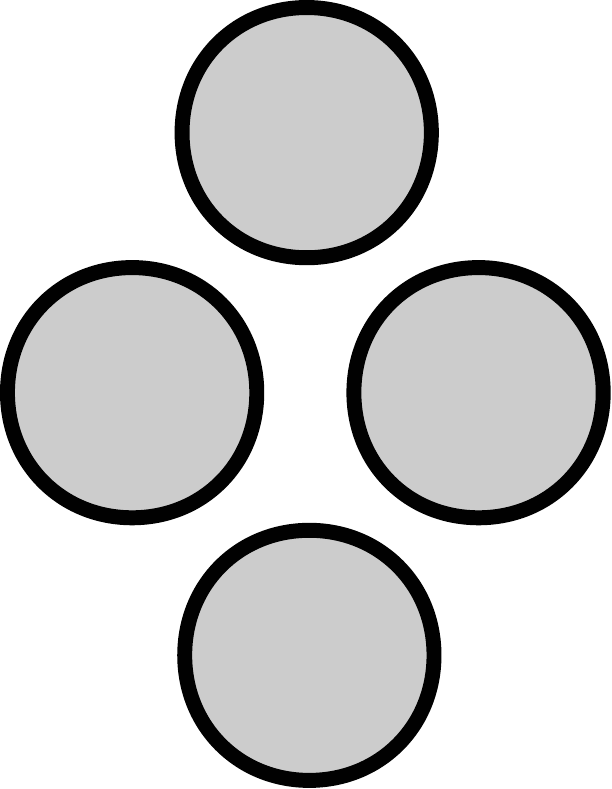} \end{tabular}  & \pmt{1/2}{0}{0}{3/4}  & 5/8 \\
\begin{tabular}{c} Heptamer \\ \includegraphics[width=0.6cm]{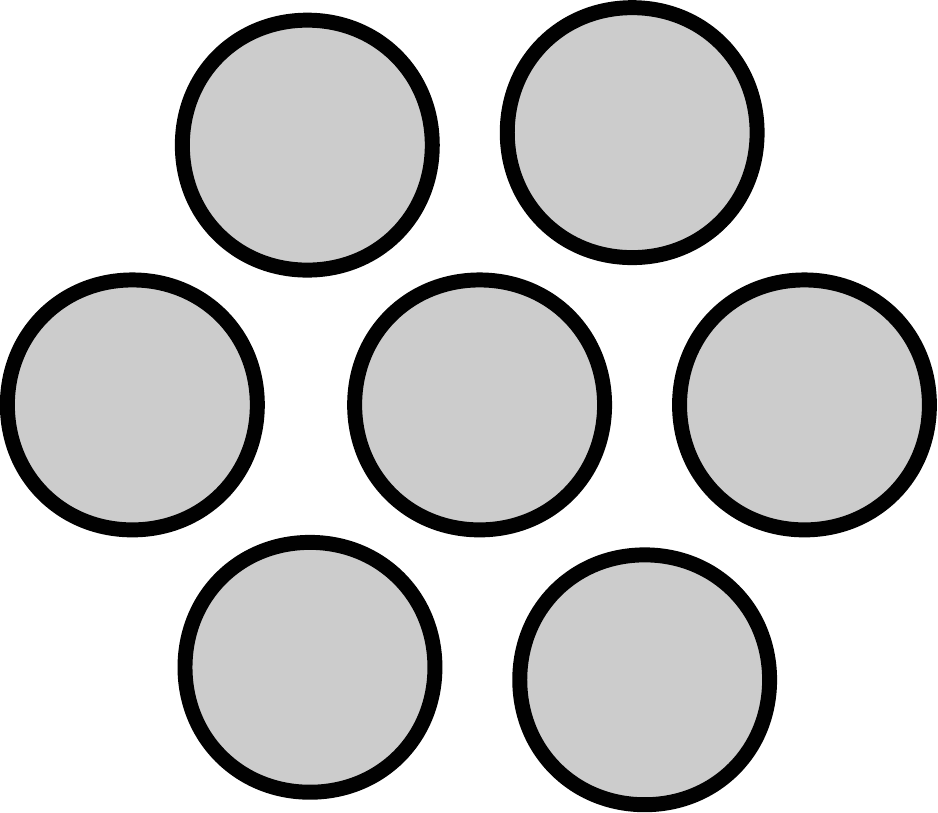} \end{tabular} & \pmt{6/7}{0}{0}{6/7}  & 6/7 \\
\begin{tabular}{c} Q-layer oligomer \\ \includegraphics[width=1.5cm]{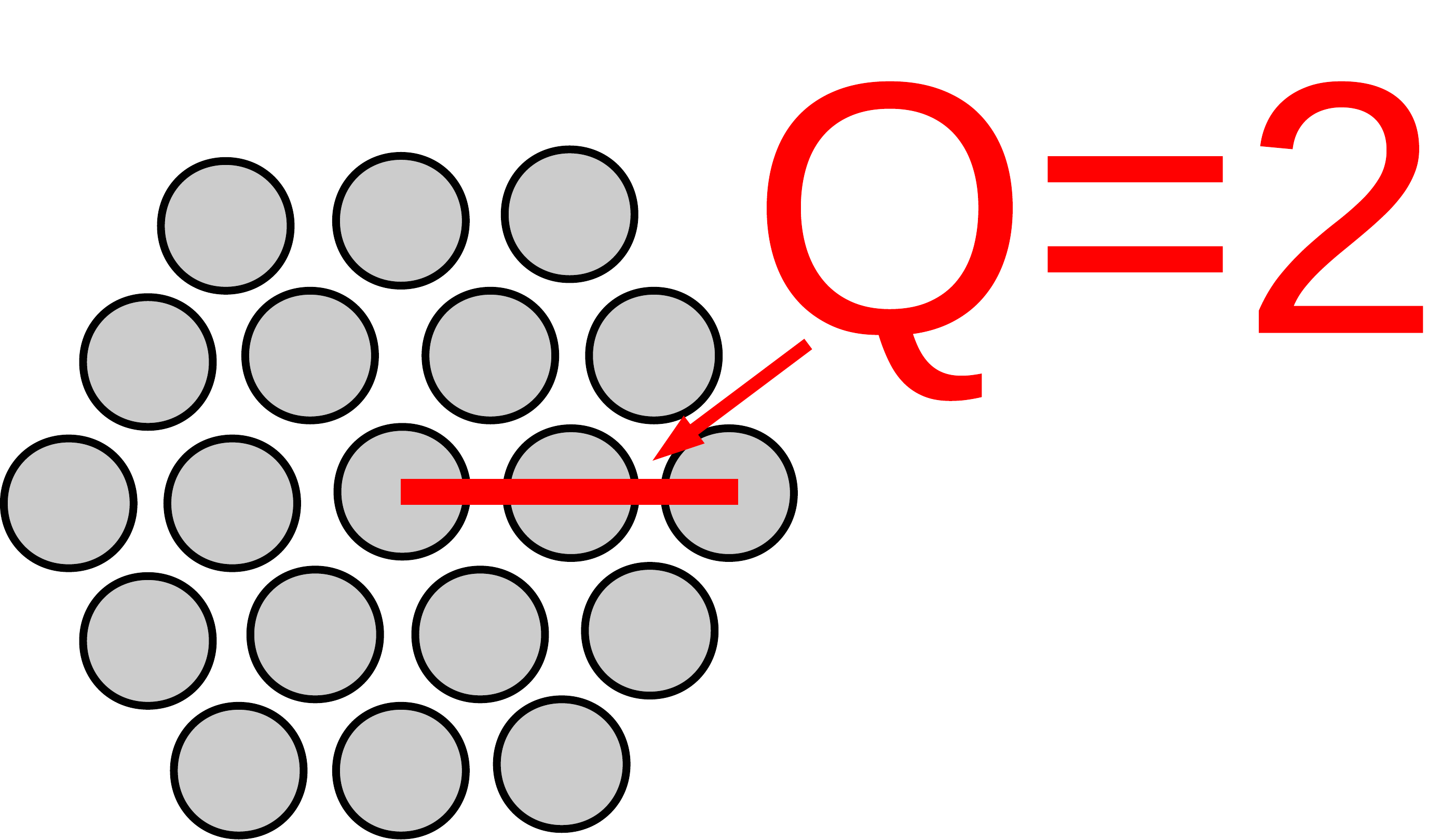} \end{tabular} & \pmt{$f(Q)$}{0}{0}{$f(Q)$}  & $f(Q) \equiv \frac{  9 Q^2 + 3 Q}{2+6 Q + 6 Q^2}$ \\
\hline
\end{tabular}

\caption{\linespread{1.0}\selectfont{}{\bf Mobility matrices $\mathcal{M}$ for several cell configurations.} For each of the configurations shown, nearest-neighbor cells have unit separation.  A $Q$-layer oligomer has $N(Q) = 1 + 3Q + 3Q^2$ cells.  $\mathcal{M}$ is given for the orientation shown in the left column; other orientations may be found by transforming the mobility tensor; $\overline{\mathcal{M}} = \frac{1}{2} \left(\mathcal{M}_{xx} + \mathcal{M}_{yy}\right)$ (see Section \ref{sec:mobrotation} below). }
\label{tab:mobs}
\end{table}

\section{Rotational transformation and averaging of the mobility matrix}
\label{sec:mobrotation}

We can compute the mobility matrix of a rotated cluster of arbitrary shape from \eq \ref{eq:matrix}.  If we rotate our cluster, which we assume is centered at the origin, by an angle $\theta$, $\left(\vb{r}^{i}\right)' = \rot(\theta) \ri$, we find that
\begin{equation}
\mathcal{M}'_{\mu\nu} = \rot_{\mu\alpha}(\theta) \mathcal{M}_{\alpha\beta} \rot_{\nu\beta}(\theta)
\end{equation}
where we have assumed the Einstein summation convention and $\rot(\theta)$ is the rotation matrix $\left(\begin{array}{cc} \cos\theta & -\sin\theta \\ \sin\theta & \cos\theta\end{array}\right)$.  In matrix terms, $\mathcal{M}' = \rot(\theta)\cdot\mathcal{M}\cdot\left[\rot(\theta)\right]^T$.  If we average over $\theta$, we find
\begin{equation}
\frac{1}{2\pi} \int_0^{2\pi} d\theta \mathcal{M}'(\theta) = \frac{1}{2} \left(
\begin{array}{cc}
\mathcal{M}_{xx}+\mathcal{M}_{yy} & \mathcal{M}_{xy}-\mathcal{M}_{yx} \\
\mathcal{M}_{yx}-\mathcal{M}_{xy} & \mathcal{M}_{xx} + \mathcal{M}_{yy} 
\end{array} \right)
\end{equation}
We can show from the definition \eq \ref{eq:matrix} that $\mathcal{M}_{\mu\nu} = \mathcal{M}_{\nu\mu}$, so the off-diagonal entries of the averaged matrix are zero, and therefore $\frac{1}{2\pi} \int_0^{2\pi} d\theta \mathcal{M}_{\mu\nu}'(\theta) = \frac{1}{2}(\mathcal{M}_{xx} + \mathcal{M}_{yy}) \delta_{\mu\nu}$.  In other words, when averaged over orientation, a cell cluster's mobility matrix is just the constant $\overline{\mathcal{M}}$ times the identity.

\section{Computing the chemotactic index}

We showed in the main paper that within our model, assuming that the cluster rearrangement is slow with respect to the polarity dynamics and thus each cell's polarity is given by a biased Ornstein-Uhlenbeck process, the velocity of a rigid cell cluster is
\begin{equation}
\vb{V} = \langle \vb{V} \rangle + \boldsymbol\Delta
\end{equation}
where $\Delta$ is a Gaussian random variable with zero mean and variance $\langle \Delta_\mu \Delta_\nu \rangle = \Gamma^2 \delta_{\mu\nu}$.  We want to compute the chemotactic index, \ci; assuming the gradient is increasing in the $x$ direction, this is
\begin{equation}
\ci = \frac{\langle V_x \rangle}{\langle | \vb{V} | \rangle}
\end{equation}
where the average is both over time and over many trajectories.  We note that this is a useful definition for us because, in our model, neither $\langle \vb{V} \rangle$ nor $\Delta$ depend on the absolute value of the chemoattractant $S$.   More care must be taken in other cases.  To compute \ci, we need to compute $\langle | \vb{V} | \rangle$.  $|\vb{V}|$ is, in our case, given by a Rice distribution, and this moment can be calculated.  
\begin{align}
\langle | \vb{V} | \rangle &= \langle \sqrt{ (\langle V_x \rangle + \Delta_x)^2 (\langle V_y\rangle + \Delta_y)^2 } \rangle \\
                           &= \frac{1}{2\pi \Gamma^2} \int d\Delta_x d \Delta_y \sqrt{ (\langle V_x \rangle + \Delta_x)^2 +(\langle V_y \rangle + \Delta_y)^2} \exp\left[ \frac{-(\Delta_x^2+\Delta_y^2)}{2 \Gamma^2}\right]\\
                           &= \frac{1}{2\pi \Gamma^2} \int dV_x dV_y V \exp \left[ -\frac{1}{2\Gamma^2}\left\{ (V_x - \langle V_x \rangle)^2 + (V_y - \langle V_y \rangle)^2 \right\} \right]
\end{align}
where $V = \sqrt{V_x^2 + V_y^2}$.  We now switch to polar coordinates, $V_x = V \cos \phi$, $V_y = V \sin \phi$, and correspondingly write $\langle V_x \rangle = \nu \cos \theta$ and $\langle V_y \rangle = \nu \sin \theta$, where $\nu^2 = \langle V_x \rangle^2 + \langle V_y \rangle^2$.  Thus, 
\begin{align}
\langle | \vb{V} | \rangle &= \frac{1}{2\pi \Gamma^2} \int_{0}^{2\pi} d\phi \int_{0}^{\infty} dV V^2 \exp \left[ -\frac{1}{2\Gamma^2}\left\{ V^2 + \nu^2 - 2 V \nu \cos (\theta-\phi) \right\} \right] \\
						   &= \frac{1}{\Gamma^2} \int_{0}^{\infty} dV V^2 \exp \left[ -\frac{1}{2\Gamma^2}\left( V^2 + \nu^2 \right) \right] I_0\left(V \nu / \Gamma^2 \right)
\end{align}
where $I_0(x)$ is the modified Bessel function of the first kind.  This integral may be evaluated, resulting in
\begin{equation}
\langle | \vb{V} | \rangle = \Gamma \sqrt{\pi/2} L_{1/2}(-\nu^2/2\Gamma^2)
\end{equation}
where the generalized Laguerre polynomial $L_{1/2}$ is given by 
\begin{equation}
L_{1/2}(x) = e^{x/2} \left[ (1-x) I_0(-x/2)-x I_1(-x/2) \right].  
\end{equation}
Within our average over trajectories, we are averaging over the orientation of the cluster; thus we expect $\langle V_y \rangle$ = 0 for a chemoattractant gradient in the $x$ direction, and $\nu = \langle V_x \rangle = \betab \tau \overline{\mathcal{M}} \partial_x S$.  $\Gamma^2 = \langle \left(V_x -\langle V_x\rangle \right)^2 \rangle = \langle \left( V_y - \langle V_y \rangle \right)^2 \rangle = \sigma^2 \tau/N$.  This leads to the result stated in the main paper, 
\begin{align}
\nonumber \ci &= \sqrt{2/\pi} c / L_{1/2}(-c^2/2)\\
c &= \frac{\langle V_x \rangle}{\sqrt{\langle \left(V_\mu-\langle V_{\mu} \rangle \right)^2 \rangle}} = \frac{\betab \tau  \overline{\mathcal{M}} \partial_x S}{\sigma \sqrt{\tau/N}}
\end{align}
where in our notation, we could also write $c = \nu/\Gamma$.  We plot the result of \eq \ref{eq:ci} in \fig \ref{fig:cic} below; we see that $\ci \to 1$ as $c \gg 1$ (corresponding to cluster velocities much larger than the noise in cluster velocity) and $\ci \to 0$ if $|c| \ll 1$ (cluster velocity much smaller than the noise).  We also note that chemotaxis could oppose the direction of the gradient (chemorepulsion) -- in this case, $\ci(-c) = -\ci(c)$.  

\begin{figure*}[ht!]
\begin{centering}
\includegraphics[width=80mm]{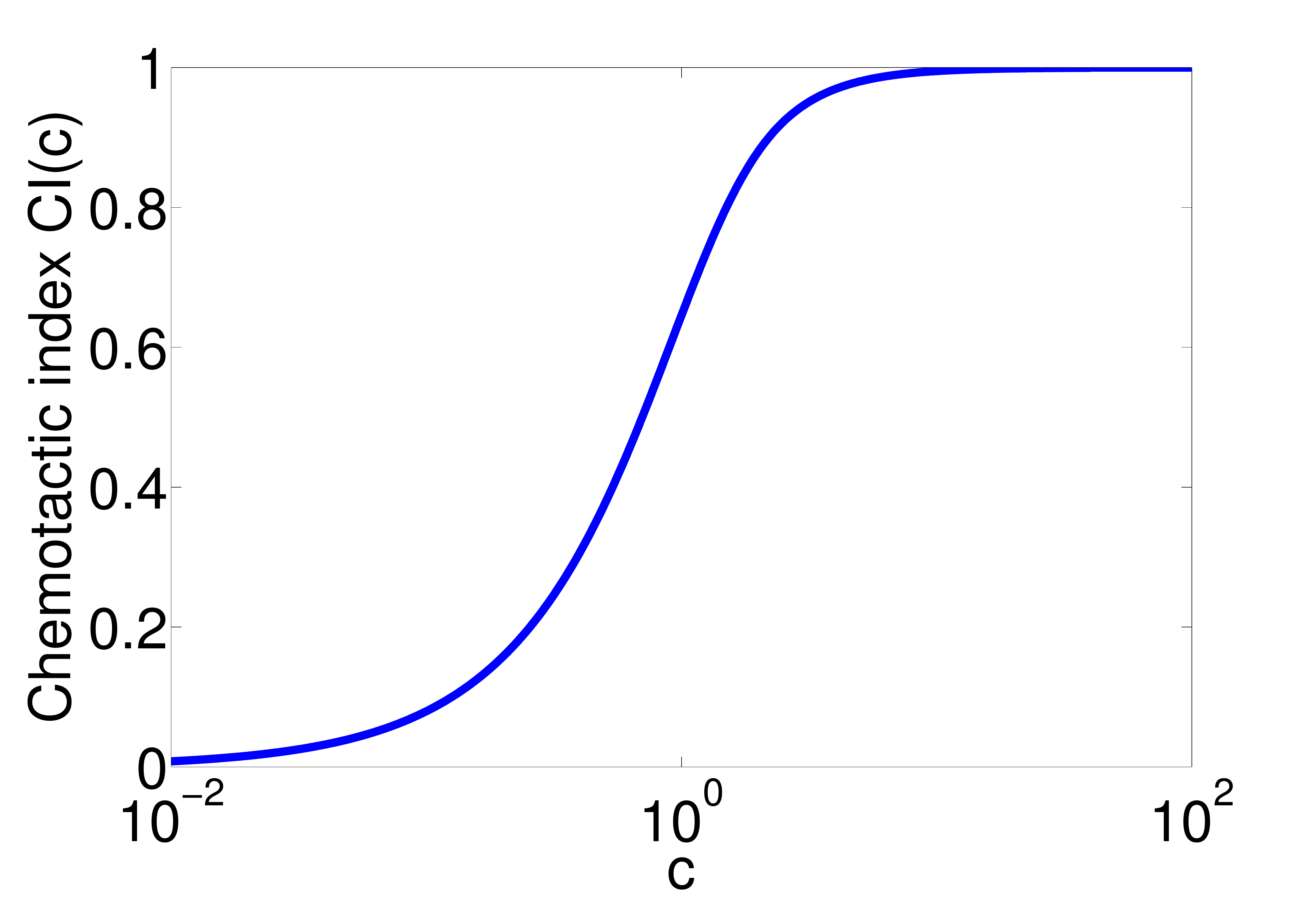}
\caption{{\bf Chemotactic index $\ci$ as a function of the parameter $c$}. }
\label{fig:cic}
\end{centering}
\end{figure*}

\section{Velocity and \ci of irregular clusters}

\begin{figure*}[ht!]
\begin{centering}
\includegraphics[width=30mm]{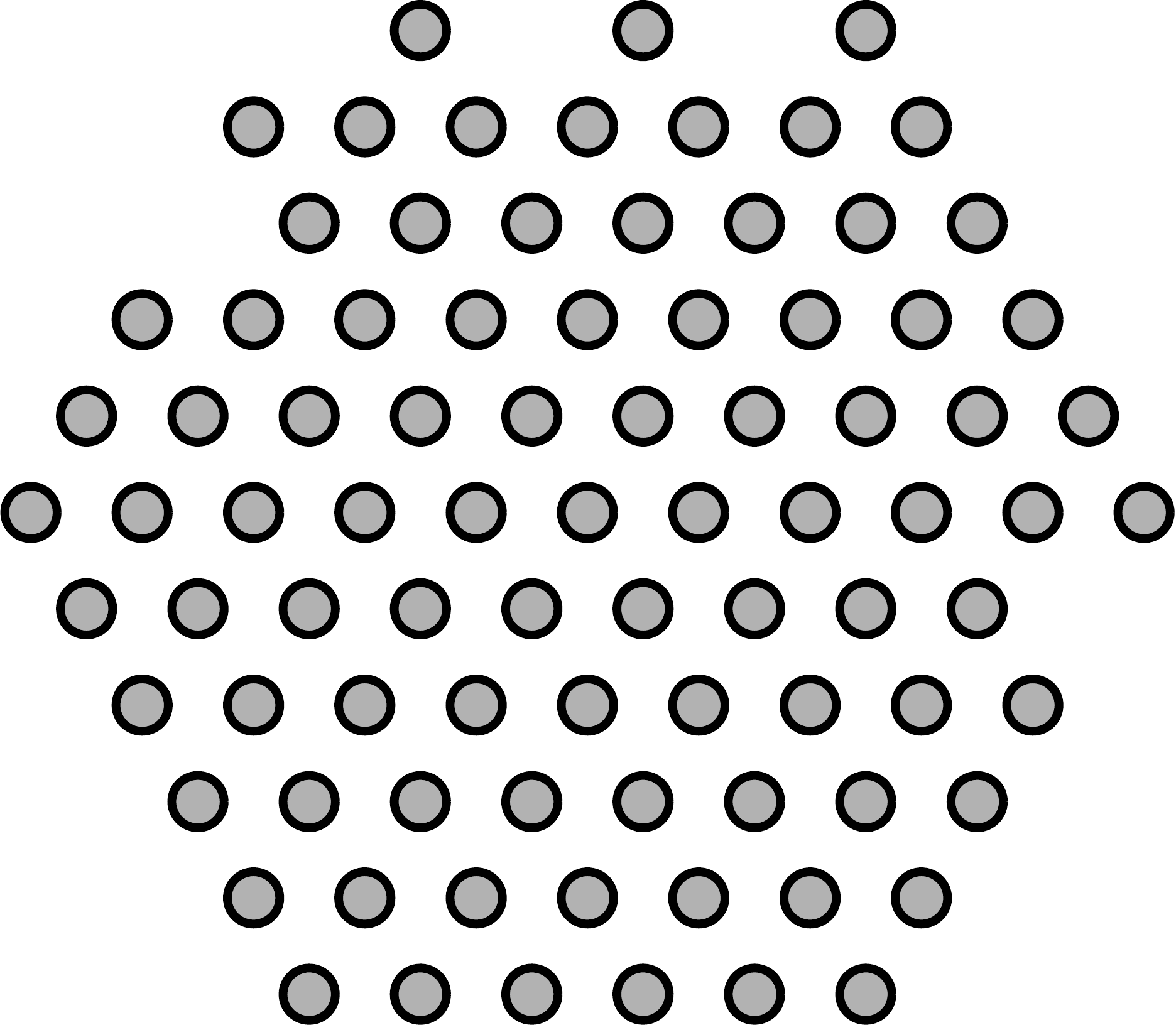}\\
\includegraphics[width=80mm]{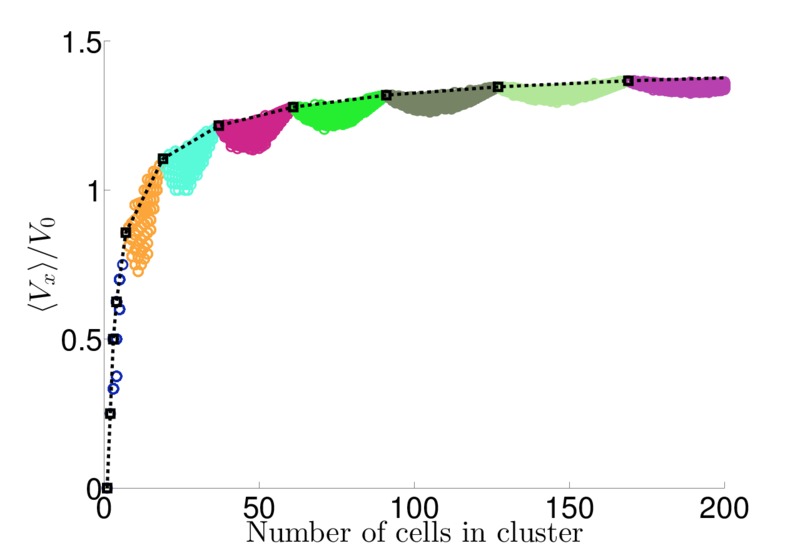}
\includegraphics[width=80mm]{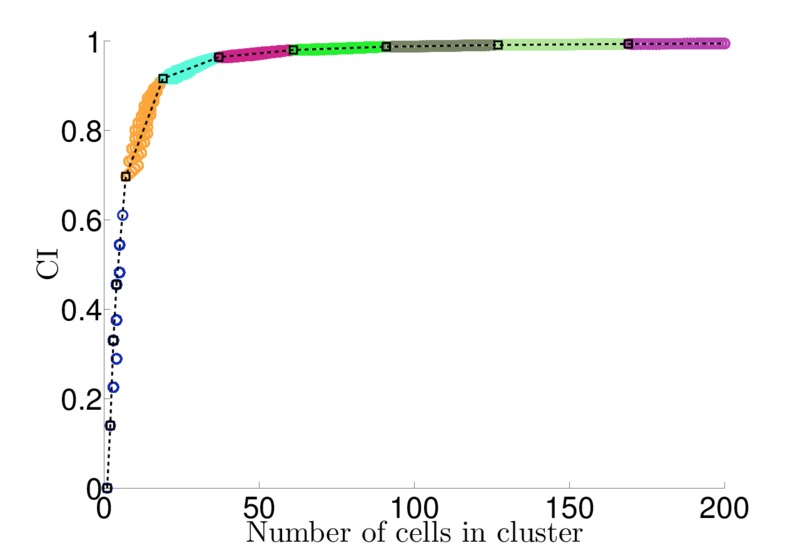}
\caption{{\bf Cluster shape effects, in addition to cell number, can affect velocity and $\ci$}. Top: illustration of $Q$-layer oligomer with a few cells removed from the external layer. Bottom: Velocity and chemotactic index for clusters of different shapes.  Different colors indicate the size of the base cluster from which cells are removed.  Black squares connected by dashed lines show the results for intact oligomers.  For the \ci plot, we apply our usual parameters and $|\nabla S| = 0.025$.   All results in this figure are theoretical results for rigid clusters only, not full simulations.}
\label{fig:shapedependence}
\end{centering}
\end{figure*}

In the main paper, we presented results on the velocity and chemotactic index of $Q$-layer oligomers.  Here, we show the velocity and chemotactic index of imperfect clusters.  We begin with a $Q$-layer oligomer, and then remove $n$ cells at random from the outer layer; this process is repeated 200 times for each $n$ from 1 to $6Q$ (the number of cells in the outer layer).  An example is presented in \fig \ref{fig:shapedependence}, with $Q = 5$ and $n = 5$ cells removed.  The mobility matrix is computed for each cluster, and used to compute $\langle V_x \rangle$ and $\ci$ (\fig \ref{fig:shapedependence}).  We see that though different configurations can lead to different mean velocities for the same number of cells, the general trend is captured by the results for intact oligomers (dashed line and square symbols in \fig \ref{fig:shapedependence}).  

\section{Transient rotation of clusters}

Though we have primarily focused on the translational motion of the cluster, rotational motion can also occur in our model, both through rotational diffusion and biased motion.  We note that transient rotational events are observed in \cite{malet2015collective}.  Under assumptions similar to our main results, clusters have mean angular velocity  $\sim\vb{A} \cdot \nabla S$, where $\vb{A}$ depends on cluster geometry.  This is again similar to an oddly-shaped particle sedimenting in a low Reynolds number flow \cite{kim2013microhydrodynamics}.  However, the symmetric clusters in Table S1 have $\vb{A} = \vb{0}$ and do not rotate.  If $\vb{A} \neq \vb{0}$, clusters rotate to a fixed angle to the gradient direction; there is no persistent rotation in a linear gradient (rotational motion may also be suppressed if $\betab$ is large).  However, in nonlinear gradients, persistent rotation of asymmetric clusters may be induced.

We can analyze potential biases by determining the net ``torque'' $L_z = \sum_i \left[\dri \times \poi \right]_z$ applied to the cluster by the cells.  This torque is, on average,
\begin{equation}
\langle L_z  \rangle = \sum_i \betab \tau S(\ri) \left[\dri \times \qi  \right]_z
\end{equation}
where $\qi = \sum_{j\sim i} \rij$ and $\dri = \ri - \vb{r}_{\textrm{cm}}$ is the displacement from the cluster center of mass.

What torque is required to cause the cluster to move at a fixed angular velocity?  For a cluster moving in a rigid rotation with angular velocity $\Omega$, the cell velocities are $\vb{v}^i = \Omega \left(-\delta r_y^i,\delta r_x^i \right)$.  To achieve this, each cell must have a polarity of $\poi = \Omega \left(-\delta r_y^i,\delta r_x^i \right)$, leading to $L_z = \Omega \sum_i |\dri|^2 $.  The angular velocity is thus related to $L_z$ by $\Omega = L_z / \sum_i |\dri|^2$. We thus find, for linear gradients, $S = S_0 + \vb{r}\cdot\nabla S$,
\begin{equation}
\langle \Omega \rangle = \betab \tau \vb{A} \cdot \nabla S \label{eq:angular}
\end{equation}
where the vector $\vb{A}$ only depends on the cluster geometry,
\begin{equation}
\vb{A} = \frac{\sum_i \dri \left[\dri \times \qi  \right]_z }{\sum_i |\dri|^2} \label{eq:A}
\end{equation}
where $\qi$ is defined as above.  (Note that $\sum_i \dri \times \qi = \sum_i \ri \times \qi = -\sum_{i, j\sim i} \frac{\ri \times\vb{r}^j}{|\ri-\vb{r}^j|} = 0$, allowing us to drop a center of mass term.)  For all of the shapes listed in Table \ref{tab:mobs}, $\vb{A} = \vb{0}$.  Cell clusters must lack an inversion symmetry to be rotated by the gradient.  

However, even if $\vb{A}\neq\vb{0}$, clusters will not persistently rotate.  We can see that if we rotate the cell cluster around its center of mass, $\vb{A}$ must also rotate as a vector.  If the gradient is along the $x$ direction, this lets us write $\langle \Omega \rangle = \langle \dot{\theta} \rangle = \betab \tau \left[ A_x(0) \cos\theta - A_y(0) \sin \theta \right]$, where $\vb{A}(0)$ is \eq \ref{eq:A} calculated for a reference geometry.  We see that if $\vb{A} \neq \vb{0}$, the cluster will rotate to a stable angle $\theta^*$ given by $\tan\theta^* = A_x(0)/A_y(0)$.  In a linear gradient, there is no persistent rotation, though in nonlinear gradients, persistent rotation of asymmetric clusters may be induced.  

We note that \eq \ref{eq:angular} is not as quantitatively accurate as the corresponding result for translational motion, at least for the parameter set in the main paper; this occurs because a small deviation from the equilibrium polarity $\langle \poi \rangle$ can create a relatively large change in torque, which often resists rotation.  For instance, if the cluster is rotated by a small angle $\delta$ without a corresponding change in $\langle \poi \rangle$, there will be a restoring torque proportional to $\betab \tau \delta$.  Thus, \eq \ref{eq:angular} will be more accurate for systems where the relaxation time $\tau$ is smaller compared to the rotational timescale of the cluster.  For similar reasons, for our rigid cluster parameter set, rotational diffusion can be quite slow (Movie S1).  

\section{Nonrigid cluster simulations}

\begin{figure*}[ht!]
\begin{centering}
\includegraphics[width=80mm]{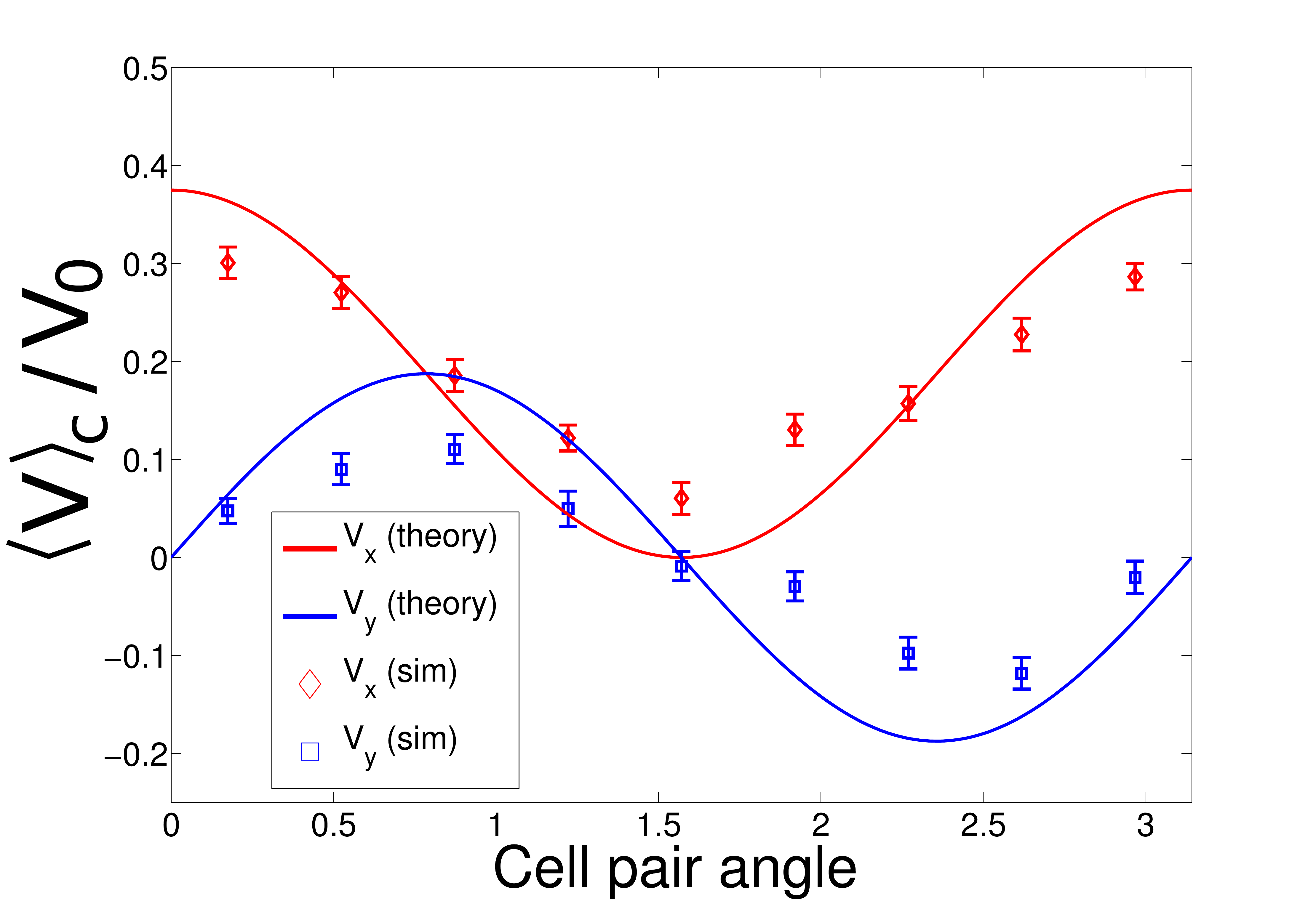}
\caption{{\bf Anisotropy exists but is lower in nonrigid clusters}. A pair of cells with our nonrigid cluster parameters has anisotropic chemotaxis, but with a slightly weaker anisotropy.  Trajectories here are measured over the time range $6.25\tau$ to $50\tau$; $n = 20,200$ trajectories are simulated. Rigid cluster results are computed using a cell-cell spacing of 0.75 to roughly match the separation seen in simulations, so the curves plotted are $0.75 \times \frac{1}{2} \cos^2\theta$ and $0.75 \times \frac{1}{2} \cos\theta\sin\theta$.}
\label{fig:pairfluid}
\end{centering}
\end{figure*}

In this section, we present additional results on nonrigid clusters.  We observe that the cluster anisotropy observed in \fig 2 of the main paper persists in fluid clusters, but is somewhat weaker (\fig \ref{fig:pairfluid}).  This occurs because the rotational diffusion of pairs of cells at our nonrigid parameter set is significantly faster than that of the rigid parameter set (compare Movie S1, Movie S2).  As the cluster's polarities $\langle \poi \rangle$ are influenced by the cluster orientation over a timescale $\tau$, as diffusion becomes faster with respect to $\tau$, anisotropy decreases.  We also note that we would not expect our rigid-cluster results to be precisely accurate even without this effect, as the pair's separation will fluctuate around its equilibrium value, changing $\mathcal{M}$.

\begin{figure*}[ht!]
\begin{centering}
\includegraphics[width=160mm]{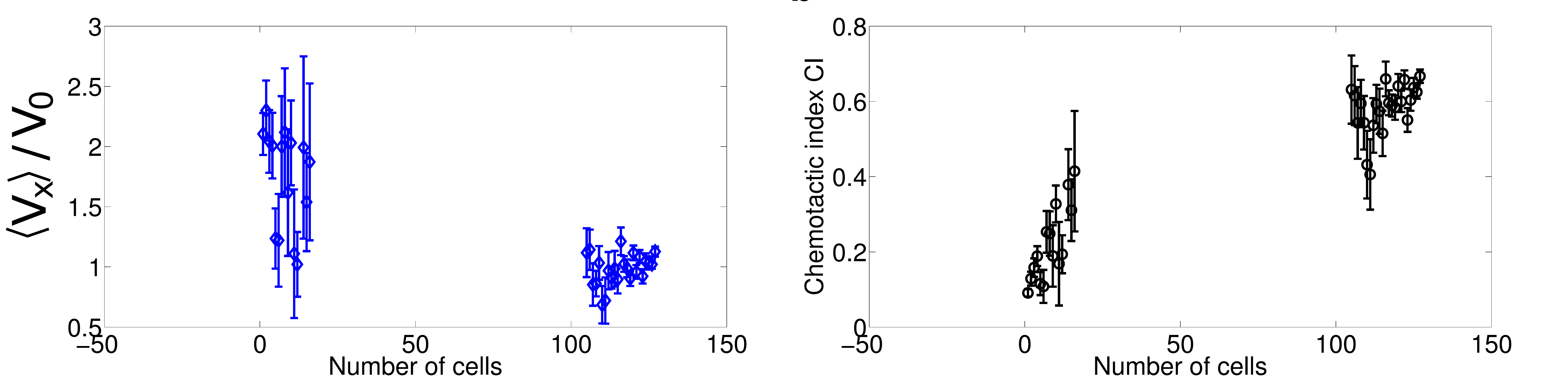}
\caption{{\bf Analyzing clusters in the process of breaking down can lead to apparent single-cell chemotaxis}.  Tracking clusters as they break off from a larger cluster leads to an increased velocity for smaller clusters, and a non-zero chemotactic index even for single cells.  This figure is generated by simulating clusters initially of 127 cells (6-layer oligomers) over a time of $50\tau$ (1000 minutes), and then computing the average velocity and average instantaneous chemotactic index $V_x/|\vb{V}|$ as a function of the number of cells in a given sub-cluster; 200 different trajectories are used.  We discard the first 25$\tau$ of the trajectory, so we are focusing on the late cluster breakup stage.  Two cells are considered to be in the same cluster if they are within a distance $D_0$ of one another (e.g. the force between them is nonzero).  We only show points in this figure where we have at least $25\tau$ of trajectory time with clusters of that size totaled over all 200 simulations. Clusters of, e.g. 50 cells are not typically seen, and therefore not shown in this figure.}
\label{fig:clusterbreakdown}
\end{centering}
\end{figure*}

We presented in the main paper a figure showing how the cluster velocity and chemotactic index depend on the cluster size (Fig. 4 of the main paper).  However, in Fig. 4, we treat all of the cells that were initially in contact as a single cluster, even if they broke apart -- thus we are plotting velocity and \ci versus the total number of cells in the simulation.  An alternate way to compute a curve showing velocity and $\ci$ as a function of cluster size would be to look at a simulation in which an initially large cluster breaks into smaller clusters (as seen in \cite{theveneau2010collective}), and then track the smaller clusters.  Doing this can yield different results, because the history of the smaller clusters matters.  In particular, clusters are more likely to break off from the side of the big cluster at higher $S$ (see e.g. Movie S2) -- leading to important biases.  For instance, we note in \fig \ref{fig:clusterbreakdown}a that smaller clusters (which have been ejected) have a larger velocity than large ones, even though isolated small clusters are slower than isolated large clusters.  Similarly, we see in \fig \ref{fig:clusterbreakdown}b that even isolated cells develop an apparent chemotactic index.  This occurs even though single isolated cells in our model have a behavior that is completely independent of the chemotactic signal -- if a single cell is isolated for a long enough time, its dynamics will again be unbiased.  

\section{Numerical details of full model simulation}

For our simulations, we solve the model equations numerically using a standard Euler-Maruyama scheme.
For our rigid cluster simulations, we adapt the cell-cell force from \cite{szabo2006phase}
\begin{equation}
\vb{F}^{ij} = \rij
\left\{
	 \begin{array}{lr}
	 v_r \left( \dij - 1 \right), & \; \; \dij < 1 \\
	 v_a \frac{ \dij-1}{D_0 - 1}, & \; \; 1 \le \dij < D_0 \\%
	 0                              & \; \; \dij > D_0
	 \end{array}
	 \right.
\end{equation}
where $\dij = |\vb{r}^i-\vb{r}^j|$.  This force is a repulsive spring below the equilibrium separation (which is one in our units, 20 microns in physical units), an attractive spring above it, and vanishes above $D_0$.  $D_0 = 1.2$ in all of our simulations, and we use $v_r = v_a = 500$.  This keeps the clusters very rigid.  

For our non-rigid cluster simulations, we adapt the many-body force chosen by Warren to simulate vapor-liquid coexistence \cite{warren2003vapor}, 
\begin{equation}
\vb{F}^{ij} = \left[A w(\dij/D_0) + B\left(\overline{\rho}^i + \overline{\rho}^j \right) w(\dij) \right] \rij  \label{eq:warren}
\end{equation}
where $w(r) = (1-r)$ for $r \le 1$ and $0$ otherwise.  The densities are defined by $\overline{\rho}^i = \sum_{j} w_\rho(\dij)$, where the sum is over all cells, including $i$, and $w_\rho(r) = \frac{6}{\pi} (1-r)^2$ for $r \le 1$ and $0$ otherwise.  The force of \eq \ref{eq:warren} is composed of an attractive force that goes to zero at a separation of $D_0$, and a repulsive force that is zero beyond the distance of cell-cell overlap (1 in our units).  Both attractive and repulsive forces have a finite value even with cells completely overlapping (``soft cores'').  The strength of the repulsive force increases with increasing cell density -- this makes the force explicitly dependent on many-body interactions.  This force makes developing fluid droplets relatively easy, even with short-range interactions \cite{warren2003vapor}.  We use $A = -23.1$ and $B = 7.35$ for our non-rigid cluster simulations unless otherwise noted.  

We initialize our clusters centered at the origin.  For rigid clusters, we start our simulations with the shapes given in Table \ref{tab:mobs} but rotated to a random angle, and a spacing of the equilibrium spacing (unity) for rigid clusters.  For non-rigid clusters, we start with the appropriate $Q$-layer cluster at a random angle, but with a spacing of $0.7$.  For non-rigid clusters, we initialize 2-, 3-, and 4-cell clusters by removing the appropriate number of cells randomly from the outer layer of a heptamer.  In both cases, we initialize the polarity $\vb{p}$ to a random value from its distribution (i.e. $\betai \tau \vb{q}^i$ plus an appropriate noise).  

\section{Table of parameters}

\begin{table}[h]
\begin{center}
\begin{tabular}{|c|c|c|}
\hline
Parameter symbol & Name & Value in our units \\
\hline
$\tau$ & Persistence time & 1 \\
$\sigma$ & Characteristic cell speed (OU noise parameter) & 1 \\ 
$\bar{\beta}$ & CIL strength & 20 for rigid clusters, 1 for nonrigid clusters\\ 
$v_a$ & Adhesion strength & 500 (rigid clusters only) \\
$v_r$ & Cell repulsion strength & 500 (rigid clusters only)\\
$D_0$ & Maximum interaction length & 1.2 \\
$S_0$ & Signal strength at origin & 1 \\
$\Delta t$ & Time step & $10^{-4}$ for rigid clusters, $5 \times 10^{-3}$ for nonrigid clusters \\
$A$ & Attraction strength for Warren force (\eq \ref{eq:warren}) & -23.1 (nonrigid clusters only) \\
$B$ & Repulsion strength for Warren force (\eq \ref{eq:warren})  & 7.35 (nonrigid clusters only) \\
\hline 
\end{tabular}
\caption{{\bf Parameters used} }
\label{tab:params}
\end{center}
\end{table}

\newpage


\begin{thebibliography}{41}
\expandafter\ifx\csname natexlab\endcsname\relax\def\natexlab#1{#1}\fi
\expandafter\ifx\csname bibnamefont\endcsname\relax
  \def\bibnamefont#1{#1}\fi
\expandafter\ifx\csname bibfnamefont\endcsname\relax
  \def\bibfnamefont#1{#1}\fi
\expandafter\ifx\csname citenamefont\endcsname\relax
  \def\citenamefont#1{#1}\fi
\expandafter\ifx\csname url\endcsname\relax
  \def\url#1{\texttt{#1}}\fi
\expandafter\ifx\csname urlprefix\endcsname\relax\def\urlprefix{URL }\fi
\providecommand{\bibinfo}[2]{#2}
\providecommand{\eprint}[2][]{\url{#2}}

\bibitem[{\citenamefont{Levine and Rappel}(2013)}]{levine2013physics}
\bibinfo{author}{\bibfnamefont{H.}~\bibnamefont{Levine}} \bibnamefont{and}
  \bibinfo{author}{\bibfnamefont{W.-J.} \bibnamefont{Rappel}},
  \bibinfo{journal}{Physics Today} \textbf{\bibinfo{volume}{66}}
  (\bibinfo{year}{2013}).

\bibitem[{\citenamefont{Segall et~al.}(1986)\citenamefont{Segall, Block, and
  Berg}}]{segall1986temporal}
\bibinfo{author}{\bibfnamefont{J.~E.} \bibnamefont{Segall}},
  \bibinfo{author}{\bibfnamefont{S.~M.} \bibnamefont{Block}}, \bibnamefont{and}
  \bibinfo{author}{\bibfnamefont{H.~C.} \bibnamefont{Berg}},
  \bibinfo{journal}{Proceedings of the National Academy of Sciences}
  \textbf{\bibinfo{volume}{83}}, \bibinfo{pages}{8987} (\bibinfo{year}{1986}).

\bibitem[{\citenamefont{Theveneau et~al.}(2010)\citenamefont{Theveneau,
  Marchant, Kuriyama, Gull, Moepps, Parsons, and
  Mayor}}]{theveneau2010collective}
\bibinfo{author}{\bibfnamefont{E.}~\bibnamefont{Theveneau}},
  \bibinfo{author}{\bibfnamefont{L.}~\bibnamefont{Marchant}},
  \bibinfo{author}{\bibfnamefont{S.}~\bibnamefont{Kuriyama}},
  \bibinfo{author}{\bibfnamefont{M.}~\bibnamefont{Gull}},
  \bibinfo{author}{\bibfnamefont{B.}~\bibnamefont{Moepps}},
  \bibinfo{author}{\bibfnamefont{M.}~\bibnamefont{Parsons}}, \bibnamefont{and}
  \bibinfo{author}{\bibfnamefont{R.}~\bibnamefont{Mayor}},
  \bibinfo{journal}{Developmental Cell} \textbf{\bibinfo{volume}{19}},
  \bibinfo{pages}{39} (\bibinfo{year}{2010}).

\bibitem[{\citenamefont{Malet-Engra et~al.}(2015)\citenamefont{Malet-Engra, Yu,
  Oldani, Rey-Barroso, Gov, Scita, and Dupr{\'e}}}]{malet2015collective}
\bibinfo{author}{\bibfnamefont{G.}~\bibnamefont{Malet-Engra}},
  \bibinfo{author}{\bibfnamefont{W.}~\bibnamefont{Yu}},
  \bibinfo{author}{\bibfnamefont{A.}~\bibnamefont{Oldani}},
  \bibinfo{author}{\bibfnamefont{J.}~\bibnamefont{Rey-Barroso}},
  \bibinfo{author}{\bibfnamefont{N.~S.} \bibnamefont{Gov}},
  \bibinfo{author}{\bibfnamefont{G.}~\bibnamefont{Scita}}, \bibnamefont{and}
  \bibinfo{author}{\bibfnamefont{L.}~\bibnamefont{Dupr{\'e}}},
  \bibinfo{journal}{Current Biology} \textbf{\bibinfo{volume}{25}},
  \bibinfo{pages}{242} (\bibinfo{year}{2015}).

\bibitem[{\citenamefont{Bianco et~al.}(2007)\citenamefont{Bianco, Poukkula,
  Cliffe, Mathieu, Luque, Fulga, and R{\o}rth}}]{bianco2007two}
\bibinfo{author}{\bibfnamefont{A.}~\bibnamefont{Bianco}},
  \bibinfo{author}{\bibfnamefont{M.}~\bibnamefont{Poukkula}},
  \bibinfo{author}{\bibfnamefont{A.}~\bibnamefont{Cliffe}},
  \bibinfo{author}{\bibfnamefont{J.}~\bibnamefont{Mathieu}},
  \bibinfo{author}{\bibfnamefont{C.~M.} \bibnamefont{Luque}},
  \bibinfo{author}{\bibfnamefont{T.~A.} \bibnamefont{Fulga}}, \bibnamefont{and}
  \bibinfo{author}{\bibfnamefont{P.}~\bibnamefont{R{\o}rth}},
  \bibinfo{journal}{Nature} \textbf{\bibinfo{volume}{448}},
  \bibinfo{pages}{362} (\bibinfo{year}{2007}).

\bibitem[{\citenamefont{R{\o}rth}(2007)}]{rorth2007collective}
\bibinfo{author}{\bibfnamefont{P.}~\bibnamefont{R{\o}rth}},
  \bibinfo{journal}{Trends in Cell Biology} \textbf{\bibinfo{volume}{17}},
  \bibinfo{pages}{575} (\bibinfo{year}{2007}).

\bibitem[{\citenamefont{Inaki et~al.}(2012)\citenamefont{Inaki, Vishnu, Cliffe,
  and R{\o}rth}}]{inaki2012effective}
\bibinfo{author}{\bibfnamefont{M.}~\bibnamefont{Inaki}},
  \bibinfo{author}{\bibfnamefont{S.}~\bibnamefont{Vishnu}},
  \bibinfo{author}{\bibfnamefont{A.}~\bibnamefont{Cliffe}}, \bibnamefont{and}
  \bibinfo{author}{\bibfnamefont{P.}~\bibnamefont{R{\o}rth}},
  \bibinfo{journal}{Proceedings of the National Academy of Sciences}
  \textbf{\bibinfo{volume}{109}}, \bibinfo{pages}{2027} (\bibinfo{year}{2012}).

\bibitem[{\citenamefont{Wang et~al.}(2010)\citenamefont{Wang, He, Wu, Hahn, and
  Montell}}]{wang2010light}
\bibinfo{author}{\bibfnamefont{X.}~\bibnamefont{Wang}},
  \bibinfo{author}{\bibfnamefont{L.}~\bibnamefont{He}},
  \bibinfo{author}{\bibfnamefont{Y.~I.} \bibnamefont{Wu}},
  \bibinfo{author}{\bibfnamefont{K.~M.} \bibnamefont{Hahn}}, \bibnamefont{and}
  \bibinfo{author}{\bibfnamefont{D.~J.} \bibnamefont{Montell}},
  \bibinfo{journal}{Nature Cell Biology} \textbf{\bibinfo{volume}{12}},
  \bibinfo{pages}{591} (\bibinfo{year}{2010}).

\bibitem[{\citenamefont{Berdahl et~al.}(2013)\citenamefont{Berdahl, Torney,
  Ioannou, Faria, and Couzin}}]{berdahl2013emergent}
\bibinfo{author}{\bibfnamefont{A.}~\bibnamefont{Berdahl}},
  \bibinfo{author}{\bibfnamefont{C.~J.} \bibnamefont{Torney}},
  \bibinfo{author}{\bibfnamefont{C.~C.} \bibnamefont{Ioannou}},
  \bibinfo{author}{\bibfnamefont{J.~J.} \bibnamefont{Faria}}, \bibnamefont{and}
  \bibinfo{author}{\bibfnamefont{I.~D.} \bibnamefont{Couzin}},
  \bibinfo{journal}{Science} \textbf{\bibinfo{volume}{339}},
  \bibinfo{pages}{574} (\bibinfo{year}{2013}).

\bibitem[{\citenamefont{Lin et~al.}(2015)\citenamefont{Lin, Yin, Wu, Inoue, and
  Levchenko}}]{lin2015interplay}
\bibinfo{author}{\bibfnamefont{B.}~\bibnamefont{Lin}},
  \bibinfo{author}{\bibfnamefont{T.}~\bibnamefont{Yin}},
  \bibinfo{author}{\bibfnamefont{Y.~I.} \bibnamefont{Wu}},
  \bibinfo{author}{\bibfnamefont{T.}~\bibnamefont{Inoue}}, \bibnamefont{and}
  \bibinfo{author}{\bibfnamefont{A.}~\bibnamefont{Levchenko}},
  \bibinfo{journal}{Nature Communications} \textbf{\bibinfo{volume}{6}}
  (\bibinfo{year}{2015}).

\bibitem[{\citenamefont{Carmona-Fontaine
  et~al.}(2008)\citenamefont{Carmona-Fontaine, Matthews, Kuriyama, Moreno,
  Dunn, Parsons, Stern, and Mayor}}]{carmona2008contact}
\bibinfo{author}{\bibfnamefont{C.}~\bibnamefont{Carmona-Fontaine}},
  \bibinfo{author}{\bibfnamefont{H.~K.} \bibnamefont{Matthews}},
  \bibinfo{author}{\bibfnamefont{S.}~\bibnamefont{Kuriyama}},
  \bibinfo{author}{\bibfnamefont{M.}~\bibnamefont{Moreno}},
  \bibinfo{author}{\bibfnamefont{G.~A.} \bibnamefont{Dunn}},
  \bibinfo{author}{\bibfnamefont{M.}~\bibnamefont{Parsons}},
  \bibinfo{author}{\bibfnamefont{C.~D.} \bibnamefont{Stern}}, \bibnamefont{and}
  \bibinfo{author}{\bibfnamefont{R.}~\bibnamefont{Mayor}},
  \bibinfo{journal}{Nature} \textbf{\bibinfo{volume}{456}},
  \bibinfo{pages}{957} (\bibinfo{year}{2008}).

\bibitem[{\citenamefont{Desai et~al.}(2013)\citenamefont{Desai, Gopal, Chen,
  and Chen}}]{desai2013contact}
\bibinfo{author}{\bibfnamefont{R.~A.} \bibnamefont{Desai}},
  \bibinfo{author}{\bibfnamefont{S.~B.} \bibnamefont{Gopal}},
  \bibinfo{author}{\bibfnamefont{S.}~\bibnamefont{Chen}}, \bibnamefont{and}
  \bibinfo{author}{\bibfnamefont{C.~S.} \bibnamefont{Chen}},
  \bibinfo{journal}{Journal of The Royal Society Interface}
  \textbf{\bibinfo{volume}{10}}, \bibinfo{pages}{20130717}
  (\bibinfo{year}{2013}).

\bibitem[{\citenamefont{Simons}(2004)}]{simons2004many}
\bibinfo{author}{\bibfnamefont{A.~M.} \bibnamefont{Simons}},
  \bibinfo{journal}{Trends in Ecology \& Evolution}
  \textbf{\bibinfo{volume}{19}}, \bibinfo{pages}{453} (\bibinfo{year}{2004}).

\bibitem[{\citenamefont{Coburn et~al.}(2013)\citenamefont{Coburn, Cerone,
  Torney, Couzin, and Neufeld}}]{coburn2013tactile}
\bibinfo{author}{\bibfnamefont{L.}~\bibnamefont{Coburn}},
  \bibinfo{author}{\bibfnamefont{L.}~\bibnamefont{Cerone}},
  \bibinfo{author}{\bibfnamefont{C.}~\bibnamefont{Torney}},
  \bibinfo{author}{\bibfnamefont{I.~D.} \bibnamefont{Couzin}},
  \bibnamefont{and} \bibinfo{author}{\bibfnamefont{Z.}~\bibnamefont{Neufeld}},
  \bibinfo{journal}{Physical Biology} \textbf{\bibinfo{volume}{10}},
  \bibinfo{pages}{046002} (\bibinfo{year}{2013}).

\bibitem[{\citenamefont{Friedl et~al.}(2012)\citenamefont{Friedl, Locker,
  Sahai, and Segall}}]{friedl2012classifying}
\bibinfo{author}{\bibfnamefont{P.}~\bibnamefont{Friedl}},
  \bibinfo{author}{\bibfnamefont{J.}~\bibnamefont{Locker}},
  \bibinfo{author}{\bibfnamefont{E.}~\bibnamefont{Sahai}}, \bibnamefont{and}
  \bibinfo{author}{\bibfnamefont{J.~E.} \bibnamefont{Segall}},
  \bibinfo{journal}{Nature Cell Biology} \textbf{\bibinfo{volume}{14}},
  \bibinfo{pages}{777} (\bibinfo{year}{2012}).

\bibitem[{\citenamefont{Aceto et~al.}(2014)\citenamefont{Aceto, Bardia,
  Miyamoto, Donaldson, Wittner, Spencer, Yu, Pely, Engstrom, Zhu
  et~al.}}]{aceto2014circulating}
\bibinfo{author}{\bibfnamefont{N.}~\bibnamefont{Aceto}},
  \bibinfo{author}{\bibfnamefont{A.}~\bibnamefont{Bardia}},
  \bibinfo{author}{\bibfnamefont{D.~T.} \bibnamefont{Miyamoto}},
  \bibinfo{author}{\bibfnamefont{M.~C.} \bibnamefont{Donaldson}},
  \bibinfo{author}{\bibfnamefont{B.~S.} \bibnamefont{Wittner}},
  \bibinfo{author}{\bibfnamefont{J.~A.} \bibnamefont{Spencer}},
  \bibinfo{author}{\bibfnamefont{M.}~\bibnamefont{Yu}},
  \bibinfo{author}{\bibfnamefont{A.}~\bibnamefont{Pely}},
  \bibinfo{author}{\bibfnamefont{A.}~\bibnamefont{Engstrom}},
  \bibinfo{author}{\bibfnamefont{H.}~\bibnamefont{Zhu}}, \bibnamefont{et~al.},
  \bibinfo{journal}{Cell} \textbf{\bibinfo{volume}{158}}, \bibinfo{pages}{1110}
  (\bibinfo{year}{2014}).

\bibitem[{\citenamefont{Selmeczi et~al.}(2005)\citenamefont{Selmeczi, Mosler,
  Hagedorn, Larsen, and Flyvbjerg}}]{selmeczi2005cell}
\bibinfo{author}{\bibfnamefont{D.}~\bibnamefont{Selmeczi}},
  \bibinfo{author}{\bibfnamefont{S.}~\bibnamefont{Mosler}},
  \bibinfo{author}{\bibfnamefont{P.~H.} \bibnamefont{Hagedorn}},
  \bibinfo{author}{\bibfnamefont{N.~B.} \bibnamefont{Larsen}},
  \bibnamefont{and}
  \bibinfo{author}{\bibfnamefont{H.}~\bibnamefont{Flyvbjerg}},
  \bibinfo{journal}{Biophysical Journal} \textbf{\bibinfo{volume}{89}},
  \bibinfo{pages}{912} (\bibinfo{year}{2005}).

\bibitem[{\citenamefont{Van~Kampen}(1992)}]{vankampen}
\bibinfo{author}{\bibfnamefont{N.~G.} \bibnamefont{Van~Kampen}},
  \emph{\bibinfo{title}{Stochastic Processes in Physics and Chemistry}},
  vol.~\bibinfo{volume}{1} (\bibinfo{publisher}{Elsevier},
  \bibinfo{year}{1992}).

\bibitem[{\citenamefont{Mayor and Carmona-Fontaine}(2010)}]{mayor2010keeping}
\bibinfo{author}{\bibfnamefont{R.}~\bibnamefont{Mayor}} \bibnamefont{and}
  \bibinfo{author}{\bibfnamefont{C.}~\bibnamefont{Carmona-Fontaine}},
  \bibinfo{journal}{Trends in Cell Biology} \textbf{\bibinfo{volume}{20}},
  \bibinfo{pages}{319} (\bibinfo{year}{2010}).

\bibitem[{\citenamefont{Camley et~al.}(2014)\citenamefont{Camley, Zhang, Zhao,
  Li, Ben-Jacob, Levine, and Rappel}}]{camley2014polarity}
\bibinfo{author}{\bibfnamefont{B.~A.} \bibnamefont{Camley}},
  \bibinfo{author}{\bibfnamefont{Y.}~\bibnamefont{Zhang}},
  \bibinfo{author}{\bibfnamefont{Y.}~\bibnamefont{Zhao}},
  \bibinfo{author}{\bibfnamefont{B.}~\bibnamefont{Li}},
  \bibinfo{author}{\bibfnamefont{E.}~\bibnamefont{Ben-Jacob}},
  \bibinfo{author}{\bibfnamefont{H.}~\bibnamefont{Levine}}, \bibnamefont{and}
  \bibinfo{author}{\bibfnamefont{W.-J.} \bibnamefont{Rappel}},
  \bibinfo{journal}{Proceedings of the National Academy of Sciences}
  \textbf{\bibinfo{volume}{111}}, \bibinfo{pages}{14770}
  (\bibinfo{year}{2014}).

\bibitem[{\citenamefont{Abercrombie}(1979)}]{abercrombie1979contact}
\bibinfo{author}{\bibfnamefont{M.}~\bibnamefont{Abercrombie}},
  \bibinfo{journal}{Nature} \textbf{\bibinfo{volume}{281}},
  \bibinfo{pages}{259} (\bibinfo{year}{1979}).

\bibitem[{\citenamefont{Kim and Karrila}(2013)}]{kim2013microhydrodynamics}
\bibinfo{author}{\bibfnamefont{S.}~\bibnamefont{Kim}} \bibnamefont{and}
  \bibinfo{author}{\bibfnamefont{S.~J.} \bibnamefont{Karrila}},
  \emph{\bibinfo{title}{Microhydrodynamics: principles and selected
  applications}} (\bibinfo{publisher}{Courier Dover Publications},
  \bibinfo{year}{2013}).

\bibitem[{\citenamefont{Han et~al.}(2006)\citenamefont{Han, Alsayed, Nobili,
  Zhang, Lubensky, and Yodh}}]{han2006brownian}
\bibinfo{author}{\bibfnamefont{Y.}~\bibnamefont{Han}},
  \bibinfo{author}{\bibfnamefont{A.}~\bibnamefont{Alsayed}},
  \bibinfo{author}{\bibfnamefont{M.}~\bibnamefont{Nobili}},
  \bibinfo{author}{\bibfnamefont{J.}~\bibnamefont{Zhang}},
  \bibinfo{author}{\bibfnamefont{T.~C.} \bibnamefont{Lubensky}},
  \bibnamefont{and} \bibinfo{author}{\bibfnamefont{A.~G.} \bibnamefont{Yodh}},
  \bibinfo{journal}{Science} \textbf{\bibinfo{volume}{314}},
  \bibinfo{pages}{626} (\bibinfo{year}{2006}).

\bibitem[{\citenamefont{Fuller et~al.}(2010)\citenamefont{Fuller, Chen, Adler,
  Groisman, Levine, Rappel, and Loomis}}]{fuller2010external}
\bibinfo{author}{\bibfnamefont{D.}~\bibnamefont{Fuller}},
  \bibinfo{author}{\bibfnamefont{W.}~\bibnamefont{Chen}},
  \bibinfo{author}{\bibfnamefont{M.}~\bibnamefont{Adler}},
  \bibinfo{author}{\bibfnamefont{A.}~\bibnamefont{Groisman}},
  \bibinfo{author}{\bibfnamefont{H.}~\bibnamefont{Levine}},
  \bibinfo{author}{\bibfnamefont{W.-J.} \bibnamefont{Rappel}},
  \bibnamefont{and} \bibinfo{author}{\bibfnamefont{W.~F.}
  \bibnamefont{Loomis}}, \bibinfo{journal}{Proceedings of the National Academy
  of Sciences} \textbf{\bibinfo{volume}{107}}, \bibinfo{pages}{9656}
  (\bibinfo{year}{2010}).

\bibitem[{\citenamefont{Angelini et~al.}(2010)\citenamefont{Angelini, Hannezo,
  Trepat, Fredberg, and Weitz}}]{angelini2010cell}
\bibinfo{author}{\bibfnamefont{T.~E.} \bibnamefont{Angelini}},
  \bibinfo{author}{\bibfnamefont{E.}~\bibnamefont{Hannezo}},
  \bibinfo{author}{\bibfnamefont{X.}~\bibnamefont{Trepat}},
  \bibinfo{author}{\bibfnamefont{J.~J.} \bibnamefont{Fredberg}},
  \bibnamefont{and} \bibinfo{author}{\bibfnamefont{D.~A.} \bibnamefont{Weitz}},
  \bibinfo{journal}{Physical Review Letters} \textbf{\bibinfo{volume}{104}},
  \bibinfo{pages}{168104} (\bibinfo{year}{2010}).

\bibitem[{\citenamefont{Angelini et~al.}(2011)\citenamefont{Angelini, Hannezo,
  Trepat, Marquez, Fredberg, and Weitz}}]{angelini2011glass}
\bibinfo{author}{\bibfnamefont{T.~E.} \bibnamefont{Angelini}},
  \bibinfo{author}{\bibfnamefont{E.}~\bibnamefont{Hannezo}},
  \bibinfo{author}{\bibfnamefont{X.}~\bibnamefont{Trepat}},
  \bibinfo{author}{\bibfnamefont{M.}~\bibnamefont{Marquez}},
  \bibinfo{author}{\bibfnamefont{J.~J.} \bibnamefont{Fredberg}},
  \bibnamefont{and} \bibinfo{author}{\bibfnamefont{D.~A.} \bibnamefont{Weitz}},
  \bibinfo{journal}{Proceedings of the National Academy of Sciences}
  \textbf{\bibinfo{volume}{108}}, \bibinfo{pages}{4714} (\bibinfo{year}{2011}).

\bibitem[{\citenamefont{Szab{\'o} et~al.}(2010)\citenamefont{Szab{\'o},
  {\"U}nnep, M{\'e}hes, Twal, Argraves, Cao, and
  Czir{\'o}k}}]{szabo2010collective}
\bibinfo{author}{\bibfnamefont{A.}~\bibnamefont{Szab{\'o}}},
  \bibinfo{author}{\bibfnamefont{R.}~\bibnamefont{{\"U}nnep}},
  \bibinfo{author}{\bibfnamefont{E.}~\bibnamefont{M{\'e}hes}},
  \bibinfo{author}{\bibfnamefont{W.}~\bibnamefont{Twal}},
  \bibinfo{author}{\bibfnamefont{W.}~\bibnamefont{Argraves}},
  \bibinfo{author}{\bibfnamefont{Y.}~\bibnamefont{Cao}}, \bibnamefont{and}
  \bibinfo{author}{\bibfnamefont{A.}~\bibnamefont{Czir{\'o}k}},
  \bibinfo{journal}{Physical Biology} \textbf{\bibinfo{volume}{7}},
  \bibinfo{pages}{046007} (\bibinfo{year}{2010}).

\bibitem[{\citenamefont{Vedula et~al.}(2013)\citenamefont{Vedula, Ravasio, Lim,
  and Ladoux}}]{vedula2013collective}
\bibinfo{author}{\bibfnamefont{S.~R.~K.} \bibnamefont{Vedula}},
  \bibinfo{author}{\bibfnamefont{A.}~\bibnamefont{Ravasio}},
  \bibinfo{author}{\bibfnamefont{C.~T.} \bibnamefont{Lim}}, \bibnamefont{and}
  \bibinfo{author}{\bibfnamefont{B.}~\bibnamefont{Ladoux}},
  \bibinfo{journal}{Physiology} \textbf{\bibinfo{volume}{28}},
  \bibinfo{pages}{370} (\bibinfo{year}{2013}).

\bibitem[{\citenamefont{Warren}(2003)}]{warren2003vapor}
\bibinfo{author}{\bibfnamefont{P.}~\bibnamefont{Warren}},
  \bibinfo{journal}{Physical Review E} \textbf{\bibinfo{volume}{68}},
  \bibinfo{pages}{066702} (\bibinfo{year}{2003}).

\bibitem[{\citenamefont{Amselem et~al.}(2012)\citenamefont{Amselem, Theves,
  Bae, Bodenschatz, and Beta}}]{amselem2012stochastic}
\bibinfo{author}{\bibfnamefont{G.}~\bibnamefont{Amselem}},
  \bibinfo{author}{\bibfnamefont{M.}~\bibnamefont{Theves}},
  \bibinfo{author}{\bibfnamefont{A.}~\bibnamefont{Bae}},
  \bibinfo{author}{\bibfnamefont{E.}~\bibnamefont{Bodenschatz}},
  \bibnamefont{and} \bibinfo{author}{\bibfnamefont{C.}~\bibnamefont{Beta}},
  \bibinfo{journal}{PloS ONE} \textbf{\bibinfo{volume}{7}},
  \bibinfo{pages}{e37213} (\bibinfo{year}{2012}).

\bibitem[{\citenamefont{Levchenko and Iglesias}(2002)}]{levchenko2002models}
\bibinfo{author}{\bibfnamefont{A.}~\bibnamefont{Levchenko}} \bibnamefont{and}
  \bibinfo{author}{\bibfnamefont{P.~A.} \bibnamefont{Iglesias}},
  \bibinfo{journal}{Biophysical Journal} \textbf{\bibinfo{volume}{82}},
  \bibinfo{pages}{50} (\bibinfo{year}{2002}).

\bibitem[{\citenamefont{Takeda et~al.}(2012)\citenamefont{Takeda, Shao, Adler,
  Charest, Loomis, Levine, Groisman, Rappel, and
  Firtel}}]{takeda2012incoherent}
\bibinfo{author}{\bibfnamefont{K.}~\bibnamefont{Takeda}},
  \bibinfo{author}{\bibfnamefont{D.}~\bibnamefont{Shao}},
  \bibinfo{author}{\bibfnamefont{M.}~\bibnamefont{Adler}},
  \bibinfo{author}{\bibfnamefont{P.~G.} \bibnamefont{Charest}},
  \bibinfo{author}{\bibfnamefont{W.~F.} \bibnamefont{Loomis}},
  \bibinfo{author}{\bibfnamefont{H.}~\bibnamefont{Levine}},
  \bibinfo{author}{\bibfnamefont{A.}~\bibnamefont{Groisman}},
  \bibinfo{author}{\bibfnamefont{W.-J.} \bibnamefont{Rappel}},
  \bibnamefont{and} \bibinfo{author}{\bibfnamefont{R.~A.}
  \bibnamefont{Firtel}}, \bibinfo{journal}{Science Signaling}
  \textbf{\bibinfo{volume}{5}}, \bibinfo{pages}{ra2} (\bibinfo{year}{2012}).

\bibitem[{\citenamefont{Sep{\'u}lveda et~al.}(2013)\citenamefont{Sep{\'u}lveda,
  Petitjean, Cochet, Grasland-Mongrain, Silberzan, and
  Hakim}}]{sepulveda2013collective}
\bibinfo{author}{\bibfnamefont{N.}~\bibnamefont{Sep{\'u}lveda}},
  \bibinfo{author}{\bibfnamefont{L.}~\bibnamefont{Petitjean}},
  \bibinfo{author}{\bibfnamefont{O.}~\bibnamefont{Cochet}},
  \bibinfo{author}{\bibfnamefont{E.}~\bibnamefont{Grasland-Mongrain}},
  \bibinfo{author}{\bibfnamefont{P.}~\bibnamefont{Silberzan}},
  \bibnamefont{and} \bibinfo{author}{\bibfnamefont{V.}~\bibnamefont{Hakim}},
  \bibinfo{journal}{PLoS Computational Biology} \textbf{\bibinfo{volume}{9}},
  \bibinfo{pages}{e1002944} (\bibinfo{year}{2013}).

\bibitem[{\citenamefont{Camley and Rappel}(2014)}]{camley2014velocity}
\bibinfo{author}{\bibfnamefont{B.~A.} \bibnamefont{Camley}} \bibnamefont{and}
  \bibinfo{author}{\bibfnamefont{W.-J.} \bibnamefont{Rappel}},
  \bibinfo{journal}{Physical Review E} \textbf{\bibinfo{volume}{89}},
  \bibinfo{pages}{062705} (\bibinfo{year}{2014}).

\bibitem[{\citenamefont{Szabo et~al.}(2006)\citenamefont{Szabo,
  Sz{\"o}ll{\"o}si, G{\"o}nci, Jur{\'a}nyi, Selmeczi, and
  Vicsek}}]{szabo2006phase}
\bibinfo{author}{\bibfnamefont{B.}~\bibnamefont{Szabo}},
  \bibinfo{author}{\bibfnamefont{G.}~\bibnamefont{Sz{\"o}ll{\"o}si}},
  \bibinfo{author}{\bibfnamefont{B.}~\bibnamefont{G{\"o}nci}},
  \bibinfo{author}{\bibfnamefont{Z.}~\bibnamefont{Jur{\'a}nyi}},
  \bibinfo{author}{\bibfnamefont{D.}~\bibnamefont{Selmeczi}}, \bibnamefont{and}
  \bibinfo{author}{\bibfnamefont{T.}~\bibnamefont{Vicsek}},
  \bibinfo{journal}{Physical Review E} \textbf{\bibinfo{volume}{74}},
  \bibinfo{pages}{061908} (\bibinfo{year}{2006}).

\bibitem[{\citenamefont{Li and Sun}(2014)}]{li2014coherent}
\bibinfo{author}{\bibfnamefont{B.}~\bibnamefont{Li}} \bibnamefont{and}
  \bibinfo{author}{\bibfnamefont{S.~X.} \bibnamefont{Sun}},
  \bibinfo{journal}{Biophysical Journal} \textbf{\bibinfo{volume}{107}},
  \bibinfo{pages}{1532} (\bibinfo{year}{2014}).

\bibitem[{\citenamefont{Czir{\'o}k et~al.}(1996)\citenamefont{Czir{\'o}k,
  Ben-Jacob, Cohen, and Vicsek}}]{czirok1996formation}
\bibinfo{author}{\bibfnamefont{A.}~\bibnamefont{Czir{\'o}k}},
  \bibinfo{author}{\bibfnamefont{E.}~\bibnamefont{Ben-Jacob}},
  \bibinfo{author}{\bibfnamefont{I.}~\bibnamefont{Cohen}}, \bibnamefont{and}
  \bibinfo{author}{\bibfnamefont{T.}~\bibnamefont{Vicsek}},
  \bibinfo{journal}{Physical Review E} \textbf{\bibinfo{volume}{54}},
  \bibinfo{pages}{1791} (\bibinfo{year}{1996}).

\bibitem[{\citenamefont{van Drongelen et~al.}(2015)\citenamefont{van Drongelen,
  Pal, Goodrich, and Idema}}]{van2014collective}
\bibinfo{author}{\bibfnamefont{R.}~\bibnamefont{van Drongelen}},
  \bibinfo{author}{\bibfnamefont{A.}~\bibnamefont{Pal}},
  \bibinfo{author}{\bibfnamefont{C.~P.} \bibnamefont{Goodrich}},
  \bibnamefont{and} \bibinfo{author}{\bibfnamefont{T.}~\bibnamefont{Idema}},
  \bibinfo{journal}{Physical Review E} \textbf{\bibinfo{volume}{91}},
  \bibinfo{pages}{032706} (\bibinfo{year}{2015}).

\bibitem[{\citenamefont{Basan et~al.}(2013)\citenamefont{Basan, Elgeti,
  Hannezo, Rappel, and Levine}}]{basan2013alignment}
\bibinfo{author}{\bibfnamefont{M.}~\bibnamefont{Basan}},
  \bibinfo{author}{\bibfnamefont{J.}~\bibnamefont{Elgeti}},
  \bibinfo{author}{\bibfnamefont{E.}~\bibnamefont{Hannezo}},
  \bibinfo{author}{\bibfnamefont{W.-J.} \bibnamefont{Rappel}},
  \bibnamefont{and} \bibinfo{author}{\bibfnamefont{H.}~\bibnamefont{Levine}},
  \bibinfo{journal}{Proceedings of the National Academy of Sciences}
  \textbf{\bibinfo{volume}{110}}, \bibinfo{pages}{2452} (\bibinfo{year}{2013}).

\bibitem[{\citenamefont{Zimmermann et~al.}(2014)\citenamefont{Zimmermann,
  Hayes, Basan, Onuchic, Rappel, and Levine}}]{zimmermann2014intercellular}
\bibinfo{author}{\bibfnamefont{J.}~\bibnamefont{Zimmermann}},
  \bibinfo{author}{\bibfnamefont{R.~L.} \bibnamefont{Hayes}},
  \bibinfo{author}{\bibfnamefont{M.}~\bibnamefont{Basan}},
  \bibinfo{author}{\bibfnamefont{J.~N.} \bibnamefont{Onuchic}},
  \bibinfo{author}{\bibfnamefont{W.-J.} \bibnamefont{Rappel}},
  \bibnamefont{and} \bibinfo{author}{\bibfnamefont{H.}~\bibnamefont{Levine}},
  \bibinfo{journal}{Biophysical Journal} \textbf{\bibinfo{volume}{107}},
  \bibinfo{pages}{548} (\bibinfo{year}{2014}).

\bibitem[{\citenamefont{Segerer et~al.}(2015)\citenamefont{Segerer,
  Th{\"u}roff, Alberola, Frey, and R{\"a}dler}}]{segerer2015emergence}
\bibinfo{author}{\bibfnamefont{F.~J.} \bibnamefont{Segerer}},
  \bibinfo{author}{\bibfnamefont{F.}~\bibnamefont{Th{\"u}roff}},
  \bibinfo{author}{\bibfnamefont{A.~P.} \bibnamefont{Alberola}},
  \bibinfo{author}{\bibfnamefont{E.}~\bibnamefont{Frey}}, \bibnamefont{and}
  \bibinfo{author}{\bibfnamefont{J.~O.} \bibnamefont{R{\"a}dler}},
  \bibinfo{journal}{Physical Review Letters} \textbf{\bibinfo{volume}{114}},
  \bibinfo{pages}{228102} (\bibinfo{year}{2015}).

\end{thebibliography}

\end{document}